\documentclass{jfm}

\usepackage{graphicx}
\usepackage{natbib}
\usepackage[usenames,dvipsnames]{xcolor}	
\usepackage{soul}
\definecolor{myorange}{HTML}{FF8700}

\ifCUPmtlplainloaded \else
  \checkfont{eurm10}
  \iffontfound
    \IfFileExists{upmath.sty}
      {\typeout{^^JFound AMS Euler Roman fonts on the system,
                   using the 'upmath' package.^^J}%
       \usepackage{upmath}}
      {\typeout{^^JFound AMS Euler Roman fonts on the system, but you
                   dont seem to have the}%
       \typeout{'upmath' package installed. JFM.cls can take advantage
                 of these fonts,^^Jif you use 'upmath' package.^^J}%
       \providecommand\upi{\pi}%
      }
  \else
    \providecommand\upi{\pi}%
  \fi
\fi

\ifCUPmtlplainloaded \else
  \checkfont{msam10}
  \iffontfound
    \IfFileExists{amssymb.sty}
      {\typeout{^^JFound AMS Symbol fonts on the system, using the
                'amssymb' package.^^J}%
       \usepackage{amssymb}%
         \let\leq=\leqslant
         \let\geq=\geqslant
      }{}
  \fi
\fi

\ifCUPmtlplainloaded \else
  \IfFileExists{amsbsy.sty}
    {\typeout{^^JFound the 'amsbsy' package on the system, using it.^^J}%
     \usepackage{amsbsy}}
    {\providecommand\boldsymbol[1]{\mbox{\boldmath $##1$}}}
\fi

\providecommand\bnabla{\boldsymbol{\nabla}}
\providecommand\bcdot{\boldsymbol{\cdot}}

\newcommand\Rey{\mbox{\textit{Re}}}  

\newsavebox{\astrutbox}
\sbox{\astrutbox}{\rule[-5pt]{0pt}{20pt}}

\newcommand\p{\ensuremath{\partial}}

\newcommand{\bv}[1]{\ensuremath{\boldsymbol{#1}}}
\newcommand{\e}[1]{\ensuremath{\times 10^{#1}}}
\newcommand{\pd}[2]{\partial_{#1}{#2}} 
\newcommand{\avg}[1]{\ensuremath{\left< #1 \right>}} 
\newcommand{\eqref}[1]{(\ref{#1})}

\title[Momentum transport in Taylor-Couette flow with vanishing curvature]{Momentum transport in Taylor-Couette flow with vanishing curvature}

\author[H. J. Brauckmann, M. Salewski and B. Eckhardt]%
{Hannes J. Brauckmann$^1$,\ns
Matthew Salewski$^{1,2}$\break
and Bruno Eckhardt$^{1,3}$%
\thanks{Email address for correspondence: bruno.eckhardt@physik.uni-marburg.de}}

\affiliation{$^1$Fachbereich Physik, Philipps-Universit\"at Marburg,
Renthof 6, D-35032 Marburg, Germany\\[\affilskip]
$^2$Department of Aerodynamics and Fluid Mechanics, Brandenburg University of Cottbus, Siemens-Halske-Ring 14, 03046 Cottbus, Germany\\[\affilskip]
$^3$J.M. Burgerscentrum, Delft University of Technology, Mekelweg 2, 2628 CD Delft, The Netherlands}

\pubyear{2010}
\volume{650}
\pagerange{119--126}
\date{\today}
\begin{document}

\maketitle

\begin{abstract}
We numerically study turbulent Taylor-Couette flow (TCF) between two independently rotating cylinders and the transition to rotating plane Couette flow (RPCF) in the limit of infinite radii. 
By using the shear Reynolds number $\Rey_S$ and rotation number $R_\Omega$ as dimensionless parameters, the transition from TCF to RPCF can be studied continuously without singularities. 
Already for radius ratios $\eta\geq0.9$ we find that the simulation results for various radius ratios and for RPCF collapse as a function of $R_\Omega$, indicating a turbulent behaviour common to both systems. 
We observe this agreement in the torque, mean momentum transport, mean profiles, and turbulent fluctuations. Moreover, the central profiles in TCF and RPCF for $R_\Omega>0$ are found to conform with inviscid neutral stability. 
Intermittent bursts that have been observed in the outer boundary layer and have been linked to the formation of a torque maximum for counter-rotation are shown to disappear as $\eta \rightarrow 1$.
The corresponding torque maximum disappears as well. Instead two new maxima of different origin appear for $\eta\geq0.9$ and RPCF, a broad and a narrow one, in contrast to the results for smaller $\eta$. 
The broad maximum at $R_\Omega=0.2$ is connected with a strong vortical flow and can be reproduced by streamwise invariant simulations. 
The narrow maximum at $R_\Omega=0.02$ only emerges with increasing $\Rey_S$ and is accompanied by an efficient and correlated momentum transport by the mean flow. Since the narrow maximum is of larger amplitude for $\Rey_S=2\e{4}$, our simulations suggest that it will dominate at even higher $\Rey_S$.
\end{abstract}


\section{Introduction}
\label{sec:intro}
\begin{figure}
  \centerline{\includegraphics[height=0.165\textheight]{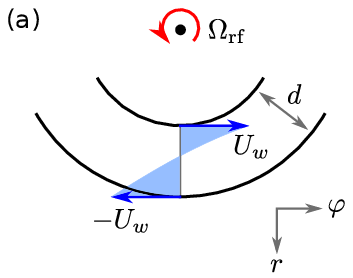}\hspace{1.5em}
	      \includegraphics[height=0.165\textheight]{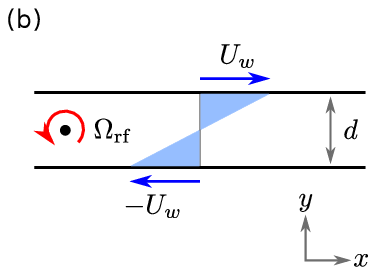}  }
  \caption{Schematic representations of the two flows. (a) TC system seen from above in the reference frame rotating with $\Omega_\mathrm{rf}$. Only one third of the azimuthal circumference is shown,
reflecting the actual computational domain for $\eta=0.5$. (b) RPCF, as obtained
from (a) for vanishing curvature, i.e. when $\eta\rightarrow 1$ for a constant gap width $d$. While the spanwise $z$-coordinate points out of the paper in both systems, the wall-normal coordinates $r$ and $y$ are antiparallel.}
\label{fig:sketch}
\end{figure}

The motion of a fluid between two independently rotating concentric cylinders, denoted as Taylor-Couette flow (TCF), serves as a fundamental model system to study the influence of rotation on turbulence. 
The two main control parameters of the system, the rotation rates of both cylinders, can be combined into two parameters that capture the key physical processes.
The differential rotation of the cylinders results in a shear gradient driving the flow, and the average of inner and outer cylinder rotation (denoted as mean system rotation) defines the magnitude of rotational influences which can either enhance or attenuate the fluid motion depending on the rotation direction. 
This decomposition into differential and mean rotation is depicted in figure \ref{fig:sketch}(a). 
Depending on the system rotation, the laminar state becomes unstable due to a centrifugal instability \citep{Rayleigh1917,Taylor1923} or without a linear instability via a subcritical transition \citep{Wendt1933,Taylor1936a,Burin2012,Deguchi2014} similar to the transition in other wall-bounded shear flows. 
Beyond the laminar regime different combinations of shear and system rotation result in a multitude of distinct flow states \citep{Coles1965,Andereck1986} that continue to shape the turbulence up to high cylinder speeds \citep{Ravelet2010,Huisman2014}.

The influence of the curvature of the cylinder walls can be captured by the ratio of the radii of the inner and outer cylinders, $\eta=r_i/r_o$, as a third control parameter. 
If the radius ratio $\eta$ falls considerably below $1$, the curvature influence on the fluid motion becomes important.
Conversely, the system curvature becomes small when $\eta$ approaches $1$ while keeping the gap width $r_o-r_i$ constant, since the cylinder radii diverge in this limit. Then, the Taylor-Couette (TC) system approximates plane Couette flow subject to spanwise system rotation, c.f. figure \ref{fig:sketch}(b), commonly called rotating plane Couette flow (RPCF). 
We here study the limiting process to vanishing curvature for turbulent TCF and its dependence on the system rotation.
For this purpose, we performed direct numerical simulations (DNS) of TCF for seven values of $\eta$ ranging between $0.5$ and $0.99$. For each radius ratio, three turbulent shear rates and various mean rotation states were realised. We complemented these TC simulations with analogous DNS of RPCF to compare directly to the curvature-free limit.

An important physical quantity in turbulent TCF is the torque needed to drive the cylinders,
which is connected to a radial transport of angular momentum by the fluid motion \citep{Marcus1984a,Eckhardt2000,Dubrulle2002}. In the turbulent regime, the torque can be characterised by its scaling with the shear strength modulated by a function that depends on the system rotation \citep{Dubrulle2005}. 
This rotation-dependence of the torque features a maximum for counter-rotating cylinders as measured and calculated for $\eta=0.72$ \citep{Paoletti2011,VanGils2011,Brauckmann2013,Ostilla2013} and for $\eta=0.5$ \citep{Merbold2013}. The torque-maximizing rotation state has been linked to the occurrence of intermittent turbulence near the outer cylinder that decreases the momentum transport efficiency \citep{VanGils2012, Brauckmann2013a} and results from the stabilisation of the outer flow region for counter-rotating cylinders. At the same time, the flow near the inner cylinder remains centrifugally unstable.
Since the curvature that causes this difference in stability between the inner and outer region 
vanishes for $\eta\rightarrow 1$, we also expect the rotation influence on the torque to change.
Evidence for such deviations in behaviour between cases with $\eta<0.9$ and larger $\eta$ have also been noted in experiments with $\eta\approx0.9$ \citep{Ravelet2010,Ostilla2014}. However, the rotation-dependence in TC systems with a radius ratio significantly larger than $0.9$ has not been explored, yet, and represents one of the main objectives of the present study.
Moreover, we will study how turbulent characteristics, such as torques, mean profiles, and fluctuations, change during the transition from low-curvature TCF to RPCF and how TCF turns into RPCF.

In the context of the transition to turbulence, the RPCF-limit of the TC system has already been studied in an approximation where certain terms in the equations of motion have been neglected \citep{Nagata1986,Nagata1990,Demay1992}. 
Nagata found that wavy vortices, which develop as an instability of streamwise invariant Taylor vortices, can be continued to non-rotating plane Couette flow where they form finite-amplitude solutions. Such wavy roll cells were also observed experimentally in RPCF at low shear rates \citep{Hiwatashi2007}.
Moreover, by considering the full equations of motion, \citet{Faisst2000} were able to continuously study the transition from TC to plane Couette flow and found that already for radius ratios $\eta$ below $1$ some characteristics of plane Couette flow can be observed. 
Furthermore, \citet{Dubrulle2005} noted that also the linear stability criterion for the laminar state in RPCF \citep{Lezius1976} is approximated by the stability boundary of TCF \citep{Esser1996} when the radius ratio $\eta$ tends to $1$.
Beyond the first bifurcation towards turbulence, further transitions create a rich set of flow states also in RPCF \citep{Tsukahara2010, Daly2014}, reminiscent of the states described by \citet{Andereck1986} in TCF. 
In an extension of these bifurcation studies at low Reynolds numbers that revealed a number of similarities between the two systems, we here investigate the continuous transition of TCF to RPCF with a focus on turbulent characteristics at high shear rates.

In the turbulent regime, the effect of rotation depends on the relative orientation of rotation and vorticity in plane Couette flow. 
Cyclonic rotation, i.e. when the rotation vector is parallel to the vorticity of the laminar base flow, was found to have a stabilising effect on the turbulence \citep{Komminaho1996} and to generate a striped pattern of coexisting laminar and turbulent regions \citep{Tsukahara2011}. Similar turbulent-laminar patterns were also observed at much higher shear rates by increasing the cyclonic rotation \citep{Brethouwer2012}. 
On the contrary, anticyclonic spanwise rotation, where the rotation is antiparallel to the base-flow vorticity, has a destabilizing effect and was found to drive vortices reminiscent of turbulent Taylor vortices \citep{Bech1997,Barri2010,Tsukahara2011}. For strong anticyclonic rotations, these vortices break up again and become disorganised. 
The influence of rotation on the strength of vortices is of substantial interest also in the present study since such vortices can enhance the momentum transport and, therefore, contribute to the formation of torque maxima as demonstrated by \citet{Brauckmann2013a}. 

Investigations of RPCF for anticyclonic rotation furthermore revealed that the mean downstream velocity profile exhibits a linear section in the centre of the gap. In this region, which can extend over a considerable
fraction of the gap, the vorticity of the mean profile compensates the imposed vorticity of the spanwise rotation \citep{Bech1997,Barri2010}, which means that the gradient of the central profile increases linearly with the imposed rotation rate. The arrangement of the profile to approximately realise zero absolute vorticity was recently also observed in a non-turbulent RPCF experiment, thus demonstrating the generality of this behaviour for unstable flows \citep{Suryadi2014}. 
A physical explanation for these central profile shapes comes from the observation that mean profiles of zero absolute vorticity comply with neutral stability in RPCF \citep{Barri2010,Suryadi2014}. Such prominent changes of the profile gradient with varying system rotation were also observed for the angular velocity in TCF \citep{Wendt1933,Ostilla2013,Ostilla2014}. 
However, it remains unclear how the mean profiles in TCF compare to those in RPCF and whether they converge in the limit of vanishing curvature ($\eta \rightarrow 1$).

The purpose of the present study is to examine the continuous transition from TCF to RPCF and to compare turbulent characteristics between both systems taking into account the following aspects. 
First, we quantify the effect of the vanishing curvature, when $\eta$ tends to unity, by the decreasing strength of intermittent fluctuations near the outer cylinder and by the local stability for counter-rotating cylinders. 
Second, we show that as these intermittent fluctuations disappear, the rotation-dependence of the torque also changes.
Third, we look for flow characteristics that prove the convergence of TCF to RPCF and seek the $\eta$ range in which the convergence can be observed.
Fourth, the magnitude of the momentum transport by coherent vortices is analysed to assess their importance for the formation of torque maxima. In addition, we evaluate the effect of turbulent fluctuations on the momentum transport.
Finally, we look for a convergence of streamwise velocity profiles and introduce a unified description of central profile gradients in TCF and RPCF.

The paper is structured as follows. In \S \ref{sec:def} we describe the unified framework in which we study TCF and RPCF, including the equations of motion, common control parameters and their ranges, the momentum transport equations and numerical aspects. The vanishing of curvature effects is studied in \S \ref{sec:burst}, followed  in \S \ref{sec:Nusselt} by an analysis of the total momentum transport as well as the vortex-induced momentum transport. 
Finally, we discuss the rotation-dependence of velocity profiles and of turbulent fluctuations contributing to the momentum transport in \S \ref{sec:charact}. We close with a summary and some concluding remarks.

\section{Definitions and numerical methods}
\label{sec:def}

Since there are several different combinations of parameters in TCF, we begin by discussing the limiting case of RPCF, so that we can then chose parameters that approach the ones in RPCF.
As usual, RPCF refers to the flow between two parallel walls that move in opposite directions and are subject to system rotation in the spanwise direction. 
We describe this system in Cartesian coordinates $(x,y,z)$ with the velocity field $\bv{u}=(u_x,u_y,u_z)$ and the walls at $y=-h$ and $y=h$ moving in the $x$-direction with velocities $-U_w$ and $U_w$, respectively. The entire system rotates with angular velocity $\bv{\Omega}=\Omega_\mathrm{rf}\bv{e}_z$ around
the spanwise axis as depicted in figure \ref{fig:sketch}(b). We chose $d=2h$ and $U_0=2U_w$ as characteristic scales for all lengths and velocities. In these units and in a reference frame rotating with $\Omega_\mathrm{rf}$, the Navier-Stokes equation and incompressibility condition become
\begin{equation}
 \pd{t}{\bv{u}} + (\bv{u}\bcdot\bnabla)\bv{u} = -\bnabla p - R_\Omega\, \bv{e}_z\times\bv{u} + \frac{1}{\Rey_S}\Delta\bv{u}, \quad \bnabla\bcdot\bv{u}=0,
 \label{eq:NS-PC}
\end{equation}
where $p$ denotes the non-dimensionalised pressure; the centrifugal force has been absorbed
into the pressure. The motion is characterised by two dimensionless parameters, the shear Reynolds number $\Rey_S$ and the rotation number $R_\Omega$ which measures the ratio between system rotation and shear,
\begin{equation}
 \Rey_S=\frac{U_0\, d}{\nu}, \quad R_\Omega=\frac{2\Omega_\mathrm{rf}\, d}{U_0},
 \label{eq:Re-PC}
\end{equation}
where $\nu$ denotes the kinematic viscosity of the fluid.
In contrast to the often used plane Couette flow Reynolds number $\Rey=U_wh/\nu$, the definition 
of the shear Reynolds numbers $\Rey_S$ is based on the full velocity difference and wall distance,  
so that $\Rey_S=4\Rey$.

TCF is the motion of a fluid between independently rotating concentric cylinders. 
Its geometric parameters are the inner and outer cylinder radii $r_i$ and $r_o$ and the axial height $L_z$. The radius ratio, defined as $\eta=r_i/r_o$, determines the curvature; it is thus the parameter that controls the transition to RPCF (which emerges for $\eta\rightarrow 1$).
The inner and outer cylinders rotate with the angular velocities $\omega_i$ and $\omega_o$, respectively, and their ratio $\mu=\omega_o/\omega_i$ or its negative, $a=-\mu$, \citep{VanGils2011,Brauckmann2013,Ostilla2013}
is often used to define the rotation state of the system. 
In order to study TCF in the same framework as RPCF, we follow the analysis of \citet{Dubrulle2005} and describe the flow in a reference frame that rotates with the angular velocity $\Omega_\mathrm{rf}$.
We make use of the cylindrical symmetry and introduce coordinates $(r,\varphi,z)$ with velocities $\bv{u}=(u_r, u_\varphi, u_z)$. In analogy to RPCF, we choose $\Omega_\mathrm{rf}$ 
such that the cylinder walls move symmetrically in the opposite direction in the rotating reference frame, i.e. such that  the condition $u_\varphi(r_i)=-u_\varphi(r_o)\equiv U_w$ is fulfilled as depicted in figure \ref{fig:sketch}(a). This gives
\begin{equation}
 r_i(\omega_i - \Omega_\mathrm{rf}) = - r_o(\omega_o-\Omega_\mathrm{rf}) 
 \quad \Leftrightarrow \quad
 \Omega_\mathrm{rf}=\frac{r_i\omega_i + r_o\omega_o}{r_i+r_o} .
 \label{eq:Omrf}
\end{equation}
Furthermore,  we choose the velocity difference between the cylinder walls as the characteristic velocity scale $U_0$ in the rotating reference frame, i.e. 
\begin{eqnarray}
 U_0 &\equiv& u_\varphi(r_i) - u_\varphi(r_o) 
 = r_i\omega_i - r_o\omega_o + (r_o-r_i)\Omega_\mathrm{rf} \nonumber \\
 &=& \frac{2}{1+\eta}\left(r_i \omega_i - \eta r_o \omega_o \right) .
 \label{eq:U0}
\end{eqnarray}
The last step in \eqref{eq:U0} results from substituting $\Omega_\mathrm{rf}$ from 
equation \eqref{eq:Omrf}. With a view towards the variables in RPCF, we take the 
gap width $d=r_o-r_i$ as the characteristic scale to measure all lengths. 
Instead of the radius ratio $\eta$, \citet{Dubrulle2005} specify a typical radius $\widetilde{r}=\sqrt{r_i r_o}$ and define the curvature number $R_C=d/\widetilde{r}$ to characterise the curvature of the system.
(The relation to $\eta$ is given in equation \eqref{eq:Re-TC} below, where the definitions of all parameters are collected.)
Using these units, the equations of motion in cylindrical coordinates \citep{Landau} transformed to the rotating reference frame can be written as
\begin{eqnarray}
 \pd{t}{\bv{u}}+(\bv{u}\bcdot\widetilde{\bnabla})\bv{u} &=& -\widetilde{\bnabla} p
      - R_\Omega\, \bv{e}_z\times\bv{u} + \frac{1}{\Rey_S}\widetilde{\Delta}\bv{u} \nonumber \\
    && + R_C\, \frac{\widetilde{r}}{r}\left[u_\varphi^2\bv{e}_r - u_r u_\varphi \bv{e}_\varphi
       +\frac{1}{\Rey_S} \left(\pd{r}{\bv{u}} + \frac{2}{r}\partial_\varphi(\bv{e}_z\times\bv{u})\right)\right] \label{eq:NS-TC} \\
    && - \frac{{R_C}^2}{\Rey_S}\, \frac{\widetilde{r}^2}{r^2}\left(u_r\bv{e}_r+u_\varphi\bv{e}_\varphi\right) \nonumber\\
    \widetilde{\bnabla}\bcdot\bv{u} &=& -R_C\, \frac{\widetilde{r}}{r} u_r
    \label{eq:cont}
\end{eqnarray}
with the modified Nabla and Laplace operators
\begin{equation}
 \widetilde{\bnabla}=\bv{e}_r \partial_r + \bv{e}_\varphi \frac{1}{r} \partial_\varphi + \bv{e}_z \partial_z, \quad
 \widetilde{\Delta} = \partial_r^2 + \frac{1}{r^2}\partial_\varphi^2 + \partial_z^2.
 \label{eq:nabla}
\end{equation}
Again the centrifugal force term is absorbed in a modified pressure $p$. 
The terms in the equations \eqref{eq:NS-TC}--\eqref{eq:nabla} are arranged in a form adapted from \citet{Faisst2000} in order to clarify that in the limit $\eta\rightarrow1$ the equations of motion \eqref{eq:NS-TC} and \eqref{eq:cont} indeed converge to the corresponding ones for RPCF \eqref{eq:NS-PC} since the additional 
terms on the right hand side vanish as $R_C\rightarrow0$. The transition between Cartesian coordinates
in RPCF and cylindrical coordinates in TC requires the identifications $x=r\varphi$ and $y\sim r$.

The equations of motion show that the TC system is characterised by three dimensionless parameters
\citep{Dubrulle2005},
\begin{eqnarray}
 \Rey_S &=& \frac{U_0\,d}{\nu} = \frac{2}{1+\eta} \left(\Rey_i - \eta\Rey_o\right), \quad
  R_\Omega = \frac{2\Omega_\mathrm{rf}\,d}{U_0}= (1-\eta) \frac{\Rey_i+\Rey_o}{\Rey_i-\eta\Rey_o}, \nonumber\\
  R_C &=& \frac{d}{\widetilde{r}} = \frac{1-\eta}{\sqrt{\eta}},
 \label{eq:Re-TC}
\end{eqnarray}
that result from \eqref{eq:Omrf} and \eqref{eq:U0}, where $Re_i=r_i \omega_i d/\nu$ and $Re_o=r_o \omega_o d/\nu$ denote the traditional Reynolds numbers that measure the dimensionless velocity of the inner and outer cylinders in the laboratory frame of reference. 
Note that $\Rey_S$ and $R_\Omega$ are defined in strict analogy to RPCF \eqref{eq:Re-PC} so that  
$R_\Omega$ has the opposite sign of the rotation number used by \citet{Dubrulle2005,Ravelet2010,Paoletti2012}. 
Finally, in these units the laminar angular velocity profile in the laboratory frame given by the circular Couette flow reads
\begin{equation}
 \omega_\mathrm{lam}(r)=\frac{1}{2}\left(R_\Omega - 1 + \frac{\widetilde{r}^2}{r^2}\right) ,
 \label{eq:omega_lam}
\end{equation}
with $r$ ranging between $r_i$ and $r_o$.

\subsection{Parameters for the limit of vanishing curvature}
We are interested in the transition from TCF to RPCF, in the limit when the radius ratio 
$\eta=r_i/r_o$ approaches one. This limit is often called the small-gap limit, since it can be obtained 
for fixed outer radius when the radius of the inner cylinder approaches that of the outer one and 
the gap width $d=r_o-r_i$ ultimately vanishes. 
However, in units of the width $d$, the radii increase like $r_i,r_o\sim1/(1-\eta)$ and diverge as $\eta$ approaches one; this is why this limit was called the case of ``large radii'' in \citet{Faisst2000}.
Moreover, the curvature vanishes ($R_C\rightarrow0$) as $\eta\rightarrow1$, suggesting that curvature effects become less important.
However, rotation effects can be maintained.
 
\begin{figure}
  \centerline{\includegraphics{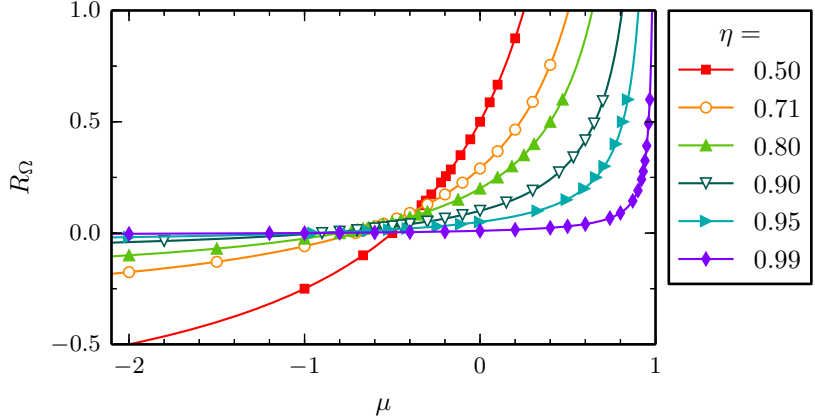}}
  \caption{Transformation between the ratio of angular velocities $\mu=\omega_o/\omega_i$ and the rotation number $R_\Omega$ for various radius ratios $\eta$. For TC systems with vanishing curvature ($\eta\rightarrow 1$), most $\mu$-values result in almost no system rotation ($R_\Omega\approx 0$) and $R_\Omega>0$ is achieved only for $\mu$ close to one. The symbols mark the values of $\mu$ and $R_\Omega$ realised by our simulations for $\Rey_S=2\e{4}$.}
\label{fig:rot-no}
\end{figure}

For every radius ratio $\eta$, the traditional Reynolds numbers can be expressed in terms of the shear Reynolds number and the rotation number as
\begin{equation}
 \Rey_i = \frac{\Rey_S}{2} + \frac{\eta}{2(1-\eta)}\Rey_S\, R_\Omega, \quad
 \Rey_o = -\frac{\Rey_S}{2} + \frac{1}{2(1-\eta)}\Rey_S\, R_\Omega
\end{equation}
and more specifically as $\Rey_i=\Rey_S/2$ and $\Rey_o=-\Rey_S/2$ in the non-rotating case ($R_\Omega=0$) for all $\eta$. In contrast, for finite rotation ($R_\Omega \ne 0$) the cylinder Reynolds numbers $\Rey_i$ and $\Rey_o$ diverge in the vanishing-curvature limit as $\eta\rightarrow1$. As a consequence, the ratio of angular velocities,
\begin{equation}
 \mu=\eta\frac{\Rey_o}{\Rey_i} = \frac{-\eta(1-\eta) + \eta R_\Omega}{(1-\eta) + \eta R_\Omega},
\end{equation}
converges to $\mu=-1$ for $R_\Omega=0$ and to $\mu=1$ for any $R_\Omega \ne 0$ in the vanishing-curvature limit, as shown in figure \ref{fig:rot-no}. These limiting values illustrate that the transition between
TCF and RPCF is singular in the traditional parameters, $\Rey_i$, $\Rey_o$, and $\mu$, but not in the parameters $\Rey_S$ and $R_\Omega$ that we will generally use here. 
We note that one exception to this parameter choice occurs in \S \ref{sec:burst} 
where the effect of curvature for counter-rotating cylinders can be characterised 
more adequately by the ratio of angular velocities $\mu$.
Furthermore, while $R_C$ measures the curvature and thus clarifies its disappearance ($R_C\rightarrow0$) for $\eta\rightarrow1$, we will identify the different TC geometries by the radius ratio $\eta$ as this parameter is more commonly used.

\subsection{Momentum transport in Taylor-Couette and plane Couette flow}
To study the mean turbulent characteristics of the flow, we calculate wall-normal profiles of the velocity and other flow fields by averaging over surfaces parallel to the wall. 
The averages are denoted  by $\avg{\cdots}$. The surfaces are concentric cylinders in the 
case of TCF, $\avg{\cdots}(r)=\avg{\cdots}_{\varphi z}$ and planes parallel to the bounding walls in the case of RPCF,  $\avg{\cdots}(y)=\avg{\cdots}_{xz}$.
Equations for the averages follow from the Navier-Stokes equation.
The average of the $\varphi$-component of the equation of motion \eqref{eq:NS-TC} of TCF results in the continuity equation for the specific angular momentum $\mathcal{L}=ru_\varphi$ 
\begin{equation}
 \partial_t \avg{\mathcal{L}} + r^{-1}\partial_r (r J_\mathcal{L})=0 \quad \mbox{with} \quad 
 J_\mathcal{L} = \avg{u_r \mathcal{L}} -\Rey_S^{-1}r^2\partial_r\avg{\omega},
 \label{eq:contTC}
\end{equation}
where $J_\mathcal{L}$ denotes the angular-momentum flux in dimensionless units \citep{Marcus1984a}, $\omega=u_\varphi/r$ the angular velocity and $r^{-1}\partial_r  (r J_\mathcal{L})$ 
the divergence in radial direction. 
In the statistically stationary state, where $\partial_t\avg{\mathcal{L}}=0$, the transverse current $J^\omega=rJ_\mathcal{L}$ 
becomes independent of the radius \citep{Eckhardt2007}.  
In physical units, the torque $T$ needed to drive the cylinders, can be calculated from the dimensionless angular-momentum flux $J_\mathcal{L}$ as \mbox{$T=\rho_f\nu^2d\,\Rey_S^2\,A(r)\,J_\mathcal{L}$}, with the dimensionless cylindrical surface area $A(r)=2\upi r L_z$ and the mass density of the fluid $\rho_f$.
The remaining prefactors compensate for the non-dimensionalization of \eqref{eq:contTC}. 
We define the dimensionless torque per axial length,
\begin{equation}
  G = \frac{T}{2\upi L_z \rho_f \nu^2 d} = Re_S^2\, r J_\mathcal{L} = Re_S^2\, J^\omega ,
  \label{eq:G}
\end{equation}
and introduce the Nusselt number,
\begin{equation}
 Nu_\omega=G/G_\mathrm{lam} =J^\omega/J^\omega_\mathrm{lam},
 \label{eq:NuTC}
\end{equation}
that measures the torque in units of the torque of the laminar profile $G_\mathrm{lam}$ 
 \citep{Dubrulle2002,Eckhardt2007a}.

Equations corresponding to \eqref{eq:contTC}--\eqref{eq:NuTC} can also be derived for RPCF. Averaging the downstream component of the equation of motion \eqref{eq:NS-PC} results in a continuity equation for the specific momentum in the $x$-direction
\begin{equation}
 \partial_t\avg{u_x} - \partial_y J^u = 0 \quad \mbox{with} \quad 
 J^u=\avg{u_y u_x} - \Rey_S^{-1}\partial_y\avg{u_x}
 \label{eq:contPC}
\end{equation}
\citep[see also][\S 7.1]{Pope}, where $J^u$ denotes the momentum flux in dimensionless units.
It is independent of the wall-normal coordinate $y$ in the statistically stationary state. 
In analogy to the torque $T$, the force $F$ needed to keep the plane walls at a constant velocity can be calculated from the momentum flux $J^u$ multiplied by the dimensionless surface area $A_{xz}=L_x L_z$ through which the flux passes, \mbox{$F=\rho_f \nu^2 \Rey_S^2 \,A_{xz} \,J^u$}. 
Finally, we define the Nusselt number for the momentum transport in RPCF as in \eqref{eq:NuTC},
\begin{equation}
 Nu_u = F/F_\mathrm{lam} = J^u/J^u_{\mathrm{lam}},
 \label{eq:NuPC}
\end{equation}
that measures the force in units of the value $F_\mathrm{lam}$ for the laminar flow.

This completes our set of relations between TCF and RPCF that we want to study here.
We have identified correspondences between 
the continuity equations \eqref{eq:contTC} and \eqref{eq:contPC}, 
the torque and the force ($T\leftrightarrow F$), 
the angular-momentum flux and the momentum flux ($J_\mathcal{L}\leftrightarrow J^u$), 
the Nusselt numbers ($Nu_\omega\leftrightarrow Nu_u$), and 
the mean profiles that enter the diffusive part of the flux in \eqref{eq:contTC} and 
\eqref{eq:contPC} \mbox{($\avg{\omega}(r)\leftrightarrow \avg{u_x}(y)$)}. 
Since we study the statistically stationary cases only, we also average in time to improve the statistics.
Using the dimensionless shear Reynolds number $\Rey_S$ and rotation number $R_\Omega$ \eqref{eq:Re-PC}, we can now study the momentum transport in both systems within the same framework and in a non-singular manner.

We conclude this section by pointing out a peculiar property of the relations that we will take up again in later sections.
While the equations of motion \eqref{eq:NS-PC} and \eqref{eq:NS-TC} depend on both driving parameters 
$\Rey_S$ and $R_\Omega$, the momentum transport equations \eqref{eq:contTC} and \eqref{eq:contPC} do not explicitly depend on the rotation number $R_\Omega$. Note that the same observation applies to RPCF \citep{Salewski2015}.
The rotation only indirectly influences the momentum fluxes $J_\mathcal{L}$ and $J^u$ 
by changing the turbulent velocity correlations ($\avg{u_r\mathcal{L}}$ and $\avg{u_y u_x}$) 
and the mean profiles ($\avg{\omega}$ and $\avg{u_x}$), as we will discuss
for the correlations in \S\S \ref{sec:LSC} and \ref{sec:fluct} and for the profiles in \S \ref{sec:slope}.

\subsection{Investigated parameter range}
\begin{figure}
  \centerline{\includegraphics{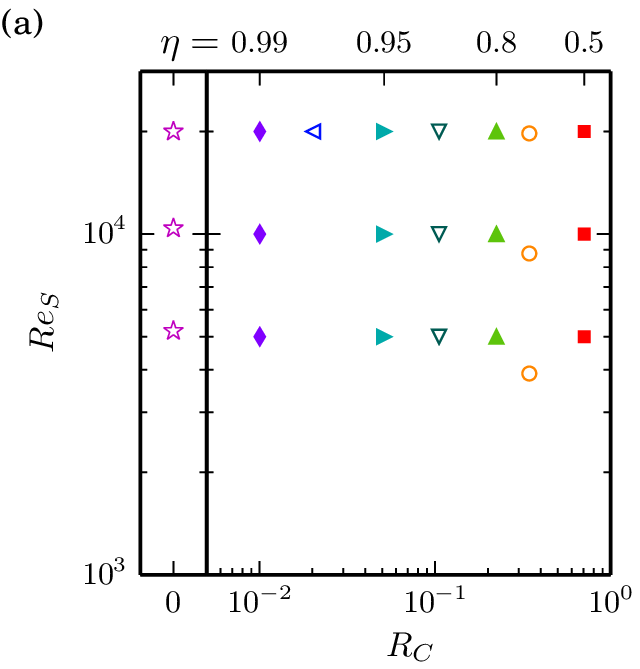}\includegraphics{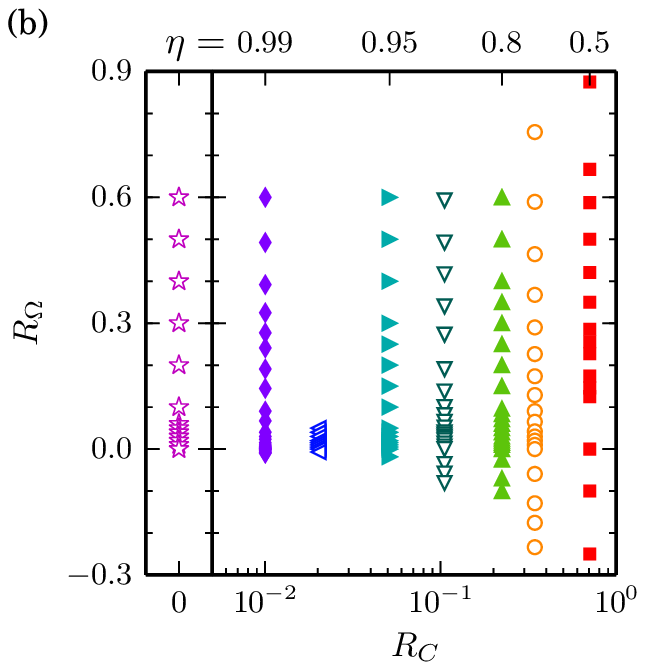}}
  \caption{Parameter space $(R_C, \Rey_S, R_\Omega)$ explored by DNS of 
  TCF to study the limit of vanishing curvature, i.e. $R_C\rightarrow 0$ equivalent to $\eta\rightarrow 1$. (a) For the shear Reynolds numbers $\Rey_S$ and curvature numbers $R_C$ indicated here, several simulations
  with $R_\Omega\in(-0.3, 0.9)$ were performed. (b) The range of rotation numbers $R_\Omega$ explored
  is shown here for  $\Rey_S=2\e{4}$ and different curvature numbers $R_C=(1-\eta)/\sqrt{\eta}$.
The curvature numbers correspond to the radius ratios 
$\eta=0.99$, $0.98$, $0.95$, $0.9$, $0.8$, $0.71$ and $0.5$. 
The extra columns for $R_C=0$ summarise our RPCF simulations.}
\label{fig:dns-range}
\end{figure}

Studies for different $\eta$ (or $R_C$) give information on when the properties of RPCF emerge from
TCF, and studies of different $Re_S$ and $R_\Omega$ give information on the
impact of shear and rotation on the turbulence. The range of parameters explored
is indicated in Fig. \ref{fig:dns-range}. They were selected as follows.

For each curvature number $R_C$ the evolution of the rotation dependence is analysed with increasing shear, by realising various mean rotation states for three shear Reynolds numbers $\Rey_S=5\e{3}$, $1\e{4}$ and $2\e{4}$. Since we adopt the results for $\eta=0.71$ from our previous study \citep{Brauckmann2013}, the simulated values of $\Rey_S=3899$, $8772$, $19737$ deviate slightly 
from the target values, as indicated in figure \ref{fig:dns-range}(a). 
Similarly, the RPCF results adopted from \citet{Salewski2015} realise $\Rey_S=5.2\e{3}$, $1.04\e{4}$ and $2\e{4}$. 
In the following, we will skip over these small differences in $\Rey_S$, as they are not significant within statistical fluctuations. 
Our highest value of $\Rey_S$ lies in the range where the transition to the fully turbulent state was 
observed in experiments for $\eta\sim0.7$ and stationary outer cylinder \citep{Lathrop1992,Lewis1999}.
We performed simulations for the range of rotation numbers $R_\Omega$ where the occurrence of a torque maximum was observed for medium radius ratios $\eta$ \citep{Dubrulle2005, Paoletti2011,VanGils2011,Paoletti2012,Brauckmann2013,Merbold2013,Ostilla2013}; this allowed us to extend these studies to $\eta$ close to one.
Moreover, rotation numbers that correspond to strong counter-rotation $\mu<0$ were selected in order to study the fate of the turbulent intermittency in the outer region \citep{Coughlin1996,VanGils2012,Brauckmann2013a}. 
Thus, the simulated rotation numbers lie mainly in the range $-0.1 \leq R_\Omega \leq 0.6$ as shown in figure \ref{fig:dns-range}(b), except for $R_C=0.02$ ($\eta=0.98$) where we focus on a narrow range 
in $R_\Omega$ only for a study of the intermittency for counter-rotation. 
The case of no system rotation,  i.e. $R_\Omega=0$, corresponds to counter-rotating 
cylinders with $\omega_o=-\eta\omega_i$.

For the highest shear Reynolds number $Re_S=2\e{4}$, the evolution of the rotation dependence is studied for seven curvature numbers $R_C>0$ (TCF) and for $R_C=0$ (RPCF), as listed in table \ref{tab:domains}. This investigation of the curvature effects compares turbulent characteristics, such as torques, mean profiles, and velocity fluctuations, between the different $R_C$ studied.

\subsection{Numerical codes and tests of convergence}
\label{sec:numeric}

\begin{table}
  \setlength{\tabcolsep}{8pt}		
  \begin{center}
  \begin{tabular}{lccccccc}
      $\eta$  & $R_C$ & $n_{\mathrm{sym}}$ & $L_x\times L_y\times L_z$  & $M_x\times M_y\times M_z$ 
      & $10^4\,\widetilde{a}_x$ & $10^4\,\widetilde{a}_y$ & $10^4\,\widetilde{a}_z$\\[3pt]
       0.5   & 0.707 &  3 & ~$\upi$\hspace{7.5pt} $\times$ 1 $\times$ 2 & ~62 $\times$ 70 $\times$ 190 & 0.33 & 0.74 & 0.23 \\
       0.71  & 0.344 &  6 & 3.09 $\times$ 1 $\times$ 2 & ~78 $\times$ 70 $\times$ 110  
       & 0.18 & 0.54 & 0.90 \\
       0.80  & 0.224 &  7 & 4.04 $\times$ 1 $\times$ 2 & ~94 $\times$ 70 $\times$ ~94 
       & 0.20 & 0.33 & 1.14 \\
       0.90  & 0.105 & 15 & 3.98 $\times$ 1 $\times$ 2 & ~94 $\times$ 70 $\times$ ~94 
       & 0.23 & 0.37 & 0.52 \\
       0.95  & 0.051 & 30 & 4.08 $\times$ 1 $\times$ 2 & ~94 $\times$ 70 $\times$ ~94 
       & 0.22 & 0.35 & 0.44 \\
       0.98  & 0.020 & 78 & 3.99 $\times$ 1 $\times$ 2 & ~94 $\times$ 70 $\times$ ~94 
       & 0.14 & 0.20 & 0.36 \\
       0.99  & 0.010 & 99 & 6.31 $\times$ 1 $\times$ 2 & 158 $\times$ 70 $\times$ ~94 
       & 0.13 & 0.30 & 0.33\\
       1     &   0   & -- & \hspace{3pt}2$\upi$\hspace{3.5pt} $\times$ 1 $\times$ 2 & 159 $\times$ 76 $\times$ ~95 & 0.42 & 0.42 & 0.38 \\
  \end{tabular}
  \caption{Size of the simulations domain at the highest $\Rey_S=2\e{4}$ for the different radius ratios $\eta$ and corresponding curvature numbers $R_C$: $L_x$, $L_y$ and $L_z$ denote the streamwise (azimuthal), wall-normal (radial) and spanwise (axial) length, and $M_x$, $M_y$, $M_z$ give the highest order of spectral modes in the corresponding direction. The azimuthal periodicity $n_{\mathrm{sym}}$ was introduced for the TC simulations to reduce the natural azimuthal length to $L_x=2\pi r_a/n_\mathrm{sym}$ with the mean radius $r_a=(r_i+r_o)/2$. The amplitudes of the highest spectral modes $\widetilde{a}_x$, $\widetilde{a}_y$ and $\widetilde{a}_z$ estimate the truncation error of the expansion in the corresponding direction. We listed the maximal amplitude values over all $R_\Omega$.}
  \label{tab:domains}
  \end{center}
\end{table}

For our investigations of TCF and RPCF we perform DNS that approximate a solution to the incompressible Navier-Stokes equation \eqref{eq:NS-TC}-\eqref{eq:cont} and \eqref{eq:NS-PC}, respectively. In these simulations we introduce a spanwise (axial) periodicity of length $L_z=2$ which is large enough to accommodate one pair of counter-rotating Taylor vortices. 
In addition, a streamwise periodicity of length $L_x=2\upi$ was assumed in the RPCF simulations while TCF is naturally periodic in this direction. However, we only simulated a domain of reduced azimuthal length $L_x$ that repeats $n_\mathrm{sym}$ times to fill the entire cylinder circumference. For $\eta=0.5$ and $\eta=0.71$ we tested that the reduced azimuthal length does not bias the computed torques for a stationary outer cylinder. The length $L_x$ was further increased for larger radius ratios, see table \ref{tab:domains}. 
As discussed in more detail in \citet{Brauckmann2013,Ostilla-Monico2015a}, the effect of these geometrical
constraints of the domain are small for single point quantities, like the torque and profiles
studied here.

For the TC simulations we employ a numerical code developed by Marc Avila and described in \citet{Meseguer2007}. To accomplish large rotation numbers $R_\Omega\sim 0.6$ for radius ratios close to $1$, both cylinders have to co-rotate rapidly and reach individual Reynolds numbers $\Rey_i$ and $\Rey_o$ of up to $6\e{5}$ while maintaining the shear $\Rey_S=2\e{4}$. Since these high absolute velocities introduce numerical instabilities, we perform these simulation cases in a reference frame rotating with the angular velocity $\Omega_\mathrm{rf}$ defined in \eqref{eq:Omrf}.
For the RPCF simulations we use \textit{Channelflow} \citep{Gibson2008,channelflow}, a pseudospectral code developed by John F. Gibson that was modified to include the Coriolis force, $-R_\Omega \bv{e}_z\times\bv{u}$.
Both numerical codes use an expansion of the velocity field in Fourier modes in the two periodic directions and in Chebyshev polynomials in the wall-normal direction, and employ a semi-implicit scheme for the time-stepping. We denote the highest order of modes employed in the streamwise, wall-normal, and spanwise direction by $M_x$, $M_y$, and $M_z$, respectively. 

The truncation error of the expansions can be assessed by the amplitudes of the highest modes $\widetilde{a}_x$, $\widetilde{a}_y$ and $\widetilde{a}_z$ which are normalised by the globally strongest mode. Our resolution tests in \citet{Brauckmann2013} revealed that these amplitudes have to drop to $\sim 10^{-4}$ in order to reach a converged torque computation. All present simulations fulfil this criterion. 
It turned out that the resolution in the axial direction could be reduced for larger $\eta$ while maintaining $\widetilde{a}_z\lesssim 10^{-4}$. This seems plausible since the boundary layer for small $\eta$ is thinner at the inner cylinder than at the outer one \citep{Eckhardt2007} and thus requires a higher resolution. Nevertheless, we did not reduce the resolution in the radial direction for large $\eta$. 
Moreover, our simulations satisfy two additional convergence criteria described by \citet{Brauckmann2013}: Agreement of the torques computed at the inner and outer cylinder to within $0.5\%$ and agreement of the energy dissipation estimated from the torque and from the volume averaged energy dissipation rate to within $1\%$.

\section{Curvature effects: Radial flow partitioning}
\label{sec:burst}
The first linear instability of TCF with counter-rotating cylinders is known to develop only in an annular 
region close to the inner cylinder while the outer region remains stable \citep{Taylor1923}. Inviscid calculations predict that the boundary between these two regions is given by the neutral surface at the radius $r_n$ where the laminar Couette profile passes through $u_\varphi=0$  \citep[pp. 276--77]{Chandrasekhar1961}.
However, already the observations by \citet{Taylor1923} as well as viscous calculations suggest that the structures from the inner partition protrude beyond $r_n$ by a factor $a(\eta)$. \citet{Esser1996} found it to be in the range between $1.4$ and $1.6$.
Beyond the first instabilities, the initially stable outer partition becomes susceptible to turbulent bursts at higher $\Rey_S\sim2\e{3}$ \citep{Coughlin1996} which give rise to a radial partitioning of the turbulent flow that persists for much higher $\Rey_S$: the coexistence of permanent turbulence in the inner partition and intermittent turbulence in the outer one was observed for $\Rey_S\sim10^6$ \citep{VanGils2012}.

In \citet{Brauckmann2013a} we argued that the intermittency in the outer partition can only occur when the extended inner unstable region does not cover the entire gap. This resulted in the prediction
\begin{equation}
 \mu_p(\eta)=-\eta^2\,\frac{(a-1)^2\eta+a^2-1}{(2a-1)\eta+1} \quad \mbox{with} \quad
 a(\eta)=(1-\eta)\left[\sqrt{\frac{(1+\eta)^3}{2(1+3\eta)}} -\eta \right]^{-1}
 \label{eq:predict}
\end{equation}
for the critical rotation ratio, $\mu_c$, so that flow partitioning occurs  for $\mu<\mu_c<0$. 
The intermittent dynamics in the outer partition comes with a reduction in momentum transport and the formation of a maximum in torque as a function of $\mu$. 
However, we already noted in our previous study that the bursting behaviour in the outer partition is a curvature effect that will not appear for $\eta\rightarrow 1$ and, consequently, \eqref{eq:predict} will become invalid in this limit. 

\begin{figure}
  \centerline{\includegraphics{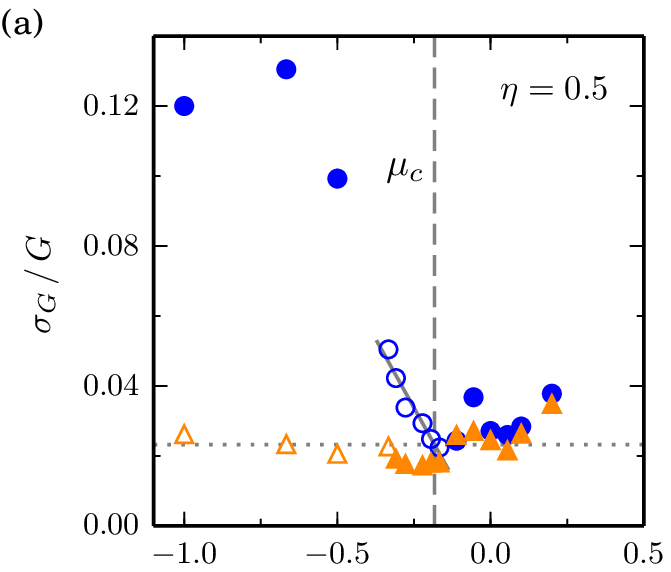}\includegraphics{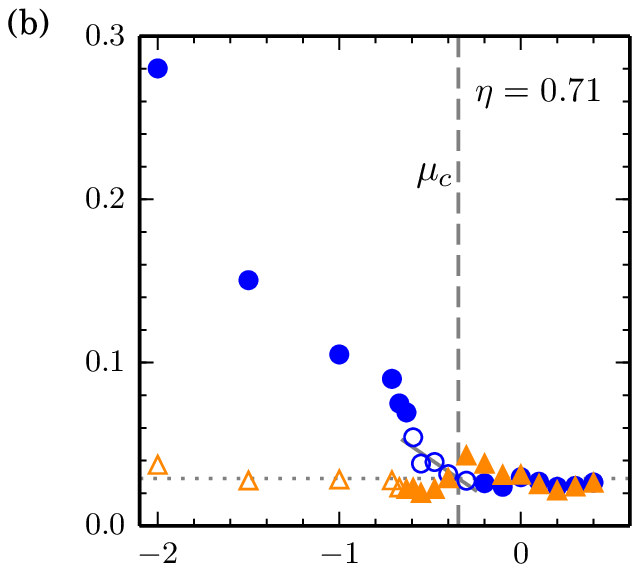}}
  \centerline{\includegraphics{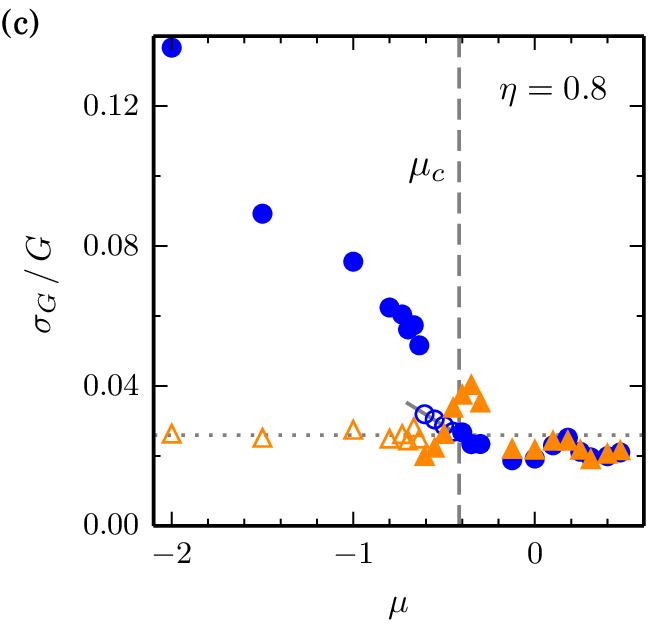}\includegraphics{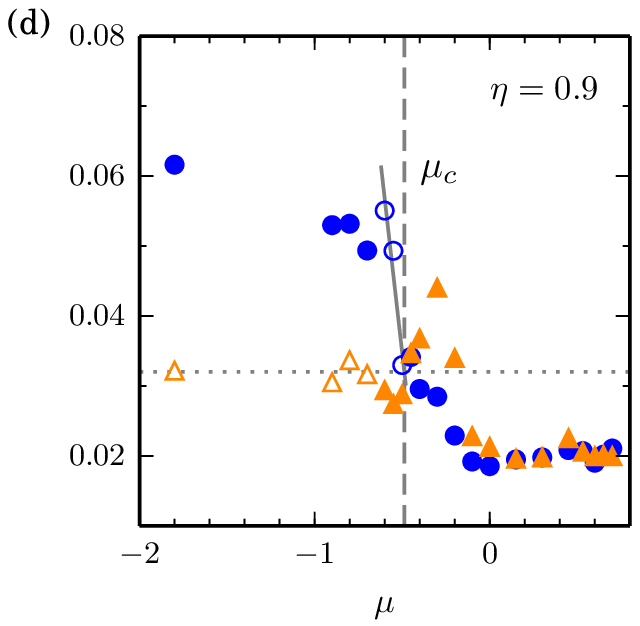}}
  \caption{Standard deviation $\sigma_G$ (divided by the mean $G$) of temporal torque fluctuations calculated at the outer cylinder ({\color{blue} \Large $\bullet\, \circ$}) and inner cylinder ({\color{myorange}  \large $\blacktriangle\, \vartriangle$}) for a constant shear $Re_S=2.0\e{4}$ and varying rotation ratio $\mu$. The dotted line marks the base level for the fluctuation at the inner cylinder for strong counter-rotation (average of open triangles). 
  The solid line is a linear fit to 6 points in (a), to 5 points in (b), to 4 points in (c) and to 3 points in (d) which are marked as open circles. 
  The vertical dashed line indicates the intersection point $\mu_c$ that defines the onset of enhanced outer fluctuations.}
\label{fig:fluct-onset}
\end{figure}

In order to study what happens to the intermittency and the torque maxima, we characterize the stronger intermittent fluctuations in the outer partition by the standard deviation $\sigma_G$ of the temporal torque fluctuations at the inner and outer cylinder for $\Rey_S=2\e{4}$. 
This previously enabled us to identify the onset of enhanced outer fluctuations as a function of $\mu$ for $\eta=0.5$ and $0.71$ \citep{Brauckmann2013a}.
We here extend the analysis to $\eta=0.8$ and $0.9$.
In the previous study, the relative fluctuation amplitude $\sigma_G/G$ was assumed to be constant (i.e. independent of $\mu$) at the inner cylinder. Therefore, the value of $\sigma_G/G$ at $\mu=0$ served as the reference value that the outer fluctuations exceed. 
However, since the data for $\eta=0.8$ and $0.9$ do not show constant inner fluctuations (figure \ref{fig:fluct-onset}c,d), we here take the average inner fluctuations ($\sigma_G/G$) for strong counter-rotation as the base level.
Figure~\ref{fig:fluct-onset} shows that the outer torque fluctuations exceed this level when $\mu$ decreases below the critical value $\mu_c(\eta)$.
Similar to the procedure used in \citet{Brauckmann2013a}, we determine the critical values
\begin{eqnarray}
\mu_c(0.5)=-0.183\pm0.014, \qquad \mu_c(0.71)=-0.344\pm0.050 \nonumber \\
\mu_c(0.8)=-0.418\pm0.025, \qquad \mu_c(0.9)\phantom{1}=-0.488\pm0.025
\end{eqnarray}
from the intersection of a linear fit to the outer fluctuations with the base level. For $\eta=0.5$ and $0.71$, the re-evaluated onsets $\mu_c$ conform with the old values within the uncertainties.
Note that the difference between the fluctuation amplitudes is less pronounced for $\eta=0.9$ than for $\eta\leq0.8$ (cf. figure \ref{fig:fluct-onset}), as expected from the weaker curvature in the first case. Moreover, the enhanced outer fluctuations due to turbulent bursting clearly differ from the situation with stationary outer cylinder or with co-rotating cylinders ($\mu\geq0$) where the fluctuation amplitudes are similar for both cylinders.

\begin{figure}
  \centerline{\includegraphics{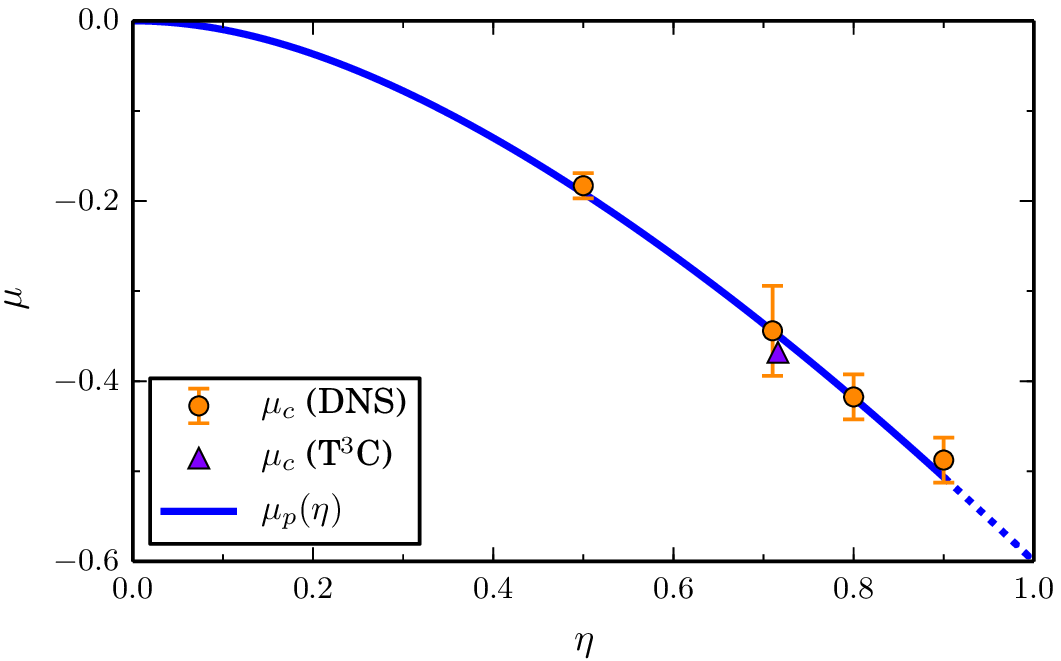}}
  \caption{Onset of enhanced outer fluctuations. The critical rotation ratio $\mu_c(\eta)$ for the onset is shown as a function of the radius ratio $\eta$. Numerical values for $Re_S=2.0\e{4}$ are extracted from figure \ref{fig:fluct-onset} and the value from the T$^3$C-experiment \citep{VanGils2012} was deduced from a bimodal distribution of angular velocities. The line indicates the prediction $\mu_p(\eta)$ from equation~\eqref{eq:predict}. As the radial partitioning disappears for $\eta \rightarrow 1$, we indicate the resulting uncertainty in the onset of enhanced fluctuations by continuing the curve with dashes for $\eta\gtrsim 0.9$.}
\label{fig:fluct-pred}
\end{figure}

Figure \ref{fig:fluct-pred} shows that the prediction from \eqref{eq:predict} agrees well with the onset of fluctuations in DNS data for $\Rey_S=2\e{4}$ (figure \ref{fig:fluct-onset}). 
Moreover, $\mu_p(\eta)$ also agrees with the critical rotation ratio for the occurrence of the bursting, detected as a bimodal distribution of angular velocities in the experiments of \citet{VanGils2012} for $\eta=0.716$. 

\begin{figure}
  \centerline{\includegraphics{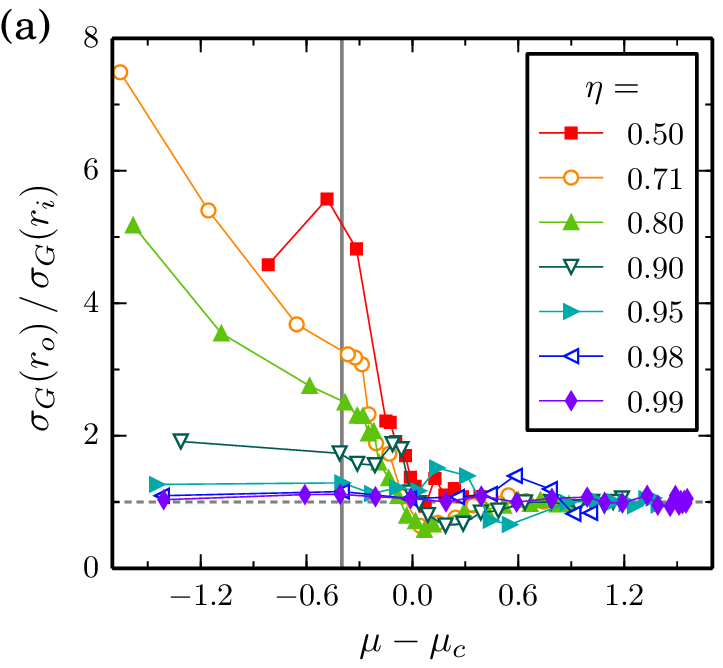}\includegraphics{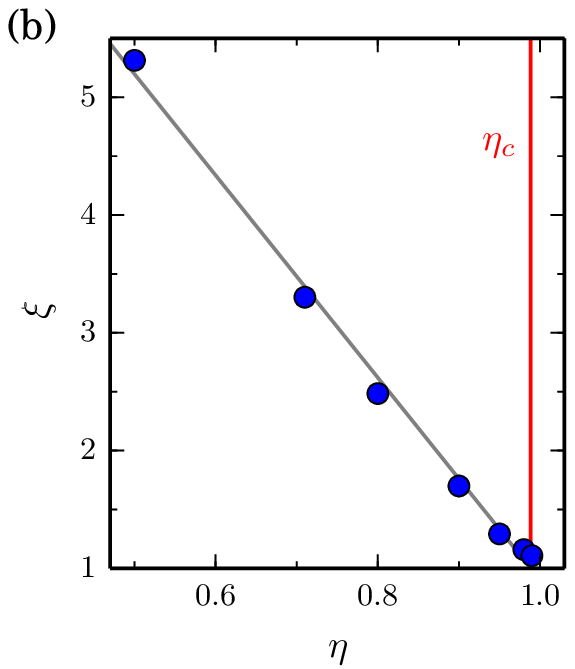}}
  \caption{Enhanced outer fluctuations for $\Rey_S=2.0\e{4}$ as a function of the radius ratio $\eta$, assessed by the standard deviation of temporal torque fluctuations $\sigma_G$. (a) Ratio of $\sigma_G$ computed at the outer and at the inner cylinder versus the shifted rotation ratio $\mu-\mu_c$, where $\mu_c$ denotes the critical value for the onset of enhanced fluctuations from figure \ref{fig:fluct-pred}. For $\eta\geq0.95$, $\mu_p$ was subtracted instead of $\mu_c$. (b) Typical fluctuation ratio $\xi=\sigma_G(r_o)/\sigma_G(r_i)$ calculated by interpolation at $\mu-\mu_c=-0.4$ marked by the vertical line in (a). The grey line represents a best fit to the data and indicates the disappearance of the fluctuation asymmetry near 
  $\eta_c=0.99\pm0.01$.}
\label{fig:fluct-crit}
\end{figure}

In figure \ref{fig:fluct-crit}(a), we extend the analysis of the torque fluctuation amplitudes to the low-curvature radius ratios $\eta=0.95$, $0.98$ and $0.99$. For the sake of visual clarity when comparing all investigated $\eta$ we show the ratio between the standard deviation $\sigma_G$ at the outer and at the inner cylinder. A value greater than one indicates enhanced outer fluctuations. 
Moreover, the rotation ratio $\mu$ is shifted by the critical value $\mu_c$ in order to align the respective regions of enhanced outer fluctuations. 
The small difference for $\eta\geq0.95$ is the reason why we omitted a similar analysis as in figure \ref{fig:fluct-onset} for the three highest radius ratios. To quantify the fluctuation asymmetry for strong counter-rotation, we define a typical fluctuation ratio $\xi=\sigma_G(r_o)/\sigma_G(r_i)$ at the shifted rotation ratio $\mu-\mu_c=-0.4$ for each radius ratio. 
Figure \ref{fig:fluct-crit}(b) shows that this asymmetry measure, $\xi$, decreases monotonically with $\eta$ towards a value of $1$ which signifies equal fluctuation strengths near the inner and outer cylinder. The fluctuation asymmetry vanishes for a radius ratio of $\eta_c=0.99\pm0.01$, which does not differ significantly from $1$. 
This suggests that small differences between inner and outer cylinder turbulence exist for all $\eta<1$. Therefore, in this case curvature has an effect for any $\eta\neq1$.

The disappearance of the fluctuation asymmetry may also be related to the 
local stability properties of the flow. The results in figure \ref{fig:fluct-crit} suggest that the partitioning into an unstable inner part and a stabilised outer part disappears as $\eta\rightarrow 1$. However, the neutral surface for counter-rotation, where the velocity profile passes through zero, discriminates the stability regions and exists for all $\eta$.
Therefore, the radial partitioning has to vanish in a different way for $\eta\rightarrow 1$, as we will see from local stability results. 
\citet{Eckhardt1995} investigated the evolution of local perturbations to the laminar Couette profile $\omega_\mathrm{lam}(r)$ along Langrangian trajectories. They calculated radially dependent eigenvalues
\begin{equation}
  \lambda_{\pm}(r) = -\frac{1}{\Rey_S}k_1^2(1+\beta) \pm 
  \left[ -\frac{2\omega_\mathrm{lam}}{r} \, \p_r\left(r^2\omega_\mathrm{lam}\right)\, \frac{\beta}{1+\beta}\right]^{1/2} 
  \quad \mbox{with} \quad \beta=\frac{k_3^2}{k_1^2}
  \label{eq:ev}
\end{equation}
for the perturbation modes with radial and axial wavenumbers $k_1$ and $k_3$ (see also
the local stability results by \citet{Dubrulle1993}). Note that the $\omega_\mathrm{lam}$-dependent factor in the second term of \eqref{eq:ev} corresponds to Rayleigh's stability discriminant. 
In the following, we focus on the largest (most unstable) eigenvalue that can be calculated for given $\Rey_S$ and $\mu$ by maximizing $\lambda_+(r)$ at the most unstable radial position $r=r_i$. For that purpose, we minimize the viscous damping (first term in \eqref{eq:ev}) by selecting the smallest possible radial wavenumber
\begin{equation}
  k_1 = \left\{
    \begin{array}{ll}
      \upi/(r_o-r_i), & \mu\geq 0 \\[2pt]
      \upi/(r_n-r_i), & \mu<0
    \end{array} \right.
    \quad \mbox{with} \quad r_n=r_i\, \sqrt{\frac{1-\mu}{\eta^2-\mu}} .
    \label{eq:k1}
\end{equation}
Subsequently, we determine the wavenumber ratio $\beta_\mathrm{max}$ that maximizes equation \eqref{eq:ev}. With these optimal values for $k_1$ and $\beta$, we study the radial variation of the largest eigenvalue $\widetilde{\lambda}_+$.

\begin{figure}
  \centerline{\includegraphics{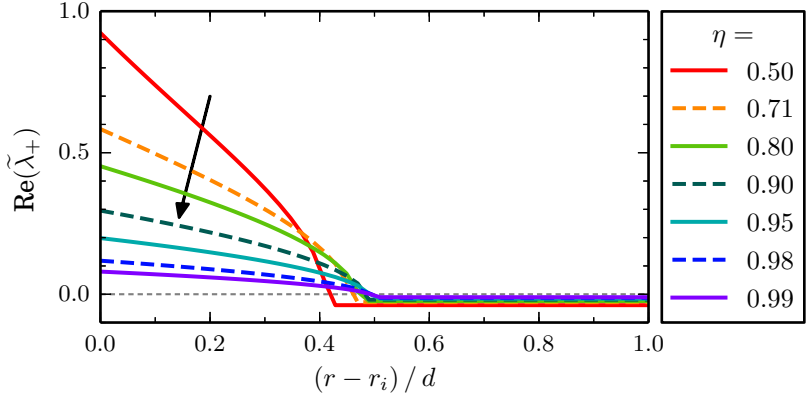}}
  \caption{Radial dependence of the centrifugal instability for $\Rey_S=2\e{4}$ and $\mu=-\eta$ (i.e. $R_\Omega=0$) where the neutral surface is located in the centre of the cylinder gap. The real part of the largest eigenvalue $\widetilde{\lambda}_+$ from the local stability analysis along Lagrangian path \citep{Eckhardt1995} is shown for various radius ratios $\eta$. The radial partitioning of stability decreases with increasing $\eta \rightarrow 1$ (indicated by the arrow).}
\label{fig:growth}
\end{figure}

To analyse the local stability for strong counter-rotation, figure \ref{fig:growth} shows radial profiles of $\widetilde{\lambda}_+$ for $\Rey_S=2\e{4}$ and the specific rotation ratio $\mu=-\eta$. This value is comparable to the value $\mu-\mu_c=-0.4$ used in the study of the fluctuation asymmetry in figure \ref{fig:fluct-crit}. 
The positive real part of the eigenvalue $\widetilde{\lambda}_+$ inside the neutral surface indicates instability in the inner partition while the outer partition is vicously damped. Most importantly, the radial variation of the local stability decreases with $\eta$ which is reminiscent of the disappearance of the radial differences in the fluctuation behaviour. 
For large $\eta$, the eigenvalues move in opposite directions: in the inner partition, they become less unstable and approach neutral stability from above, whereas in the outer partition, they approach the neutral value from below.

The results of this section show that the radial partitioning of the flow that gives rise to enhanced outer fluctuations conforms with the prediction \eqref{eq:predict} for $\eta\leq 0.9$. However, this curvature effect as well as the radial variation of the local stability disappear with vanishing curvature for 
$\eta\rightarrow 1$. The disappearance of the intermittent bursts in the outer partition is 
also relevant for an understanding of the variation of torque, since the 
appearance of these bursts has been linked to the emergence of a torque maximum \citep{VanGils2012,Brauckmann2013a}. The absence of this intermittency should therefore change the variation of torque with rotation, as we now discuss.

\section{Variation of the momentum transport}
\label{sec:Nusselt}

\begin{figure}
  \centerline{\includegraphics{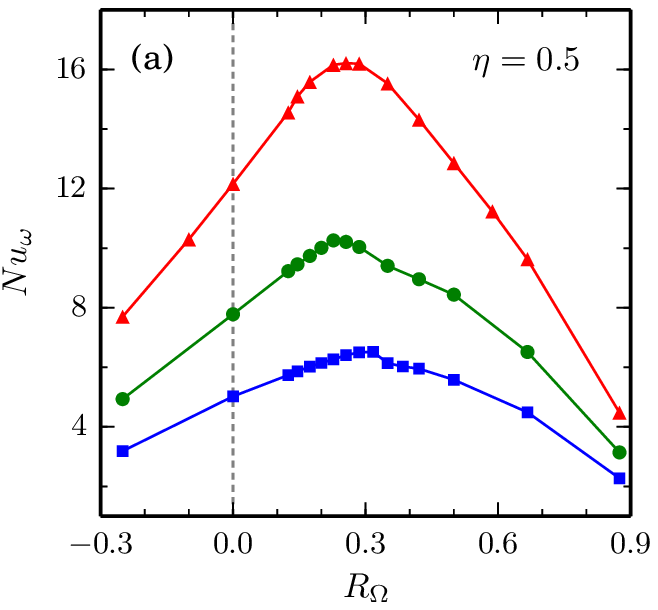}\includegraphics{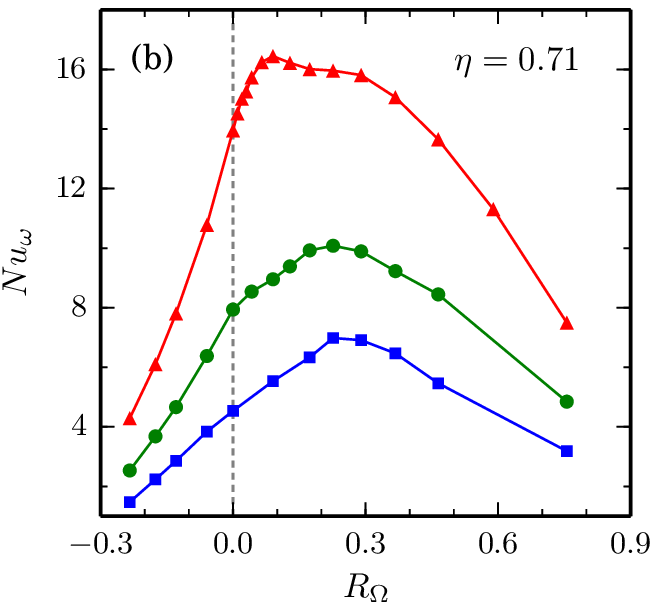}}
  \centerline{\includegraphics{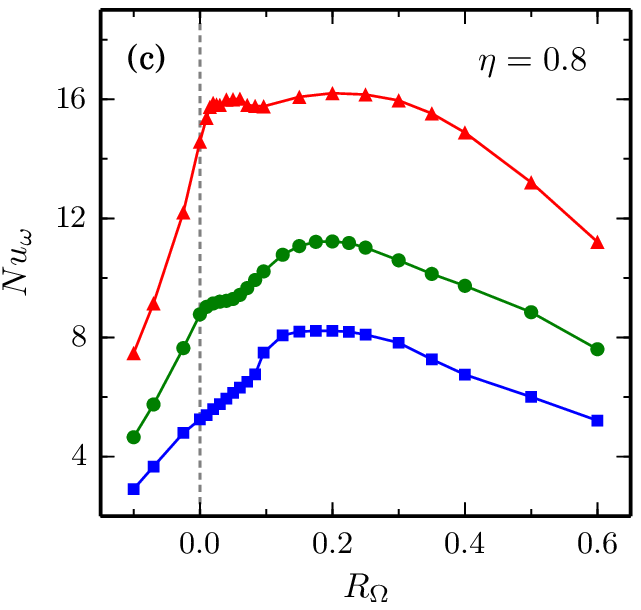}\includegraphics{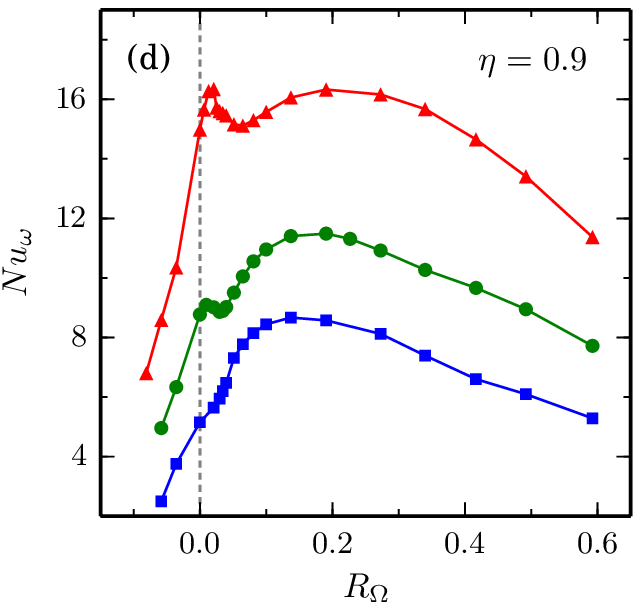}}
  \centerline{\includegraphics{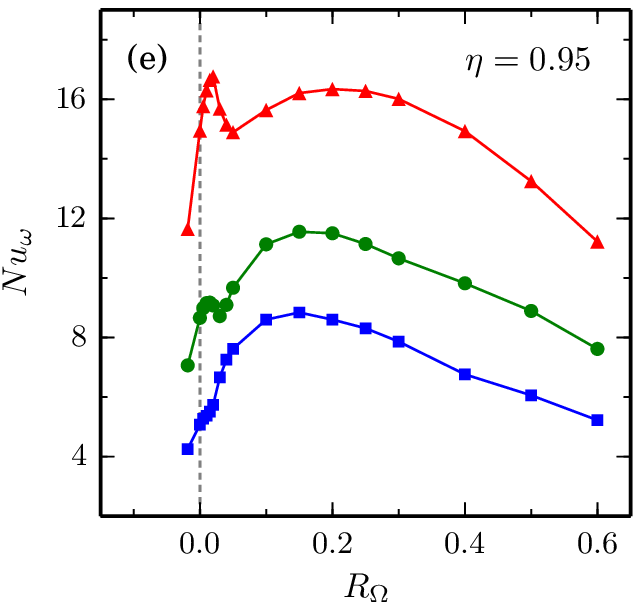}\includegraphics{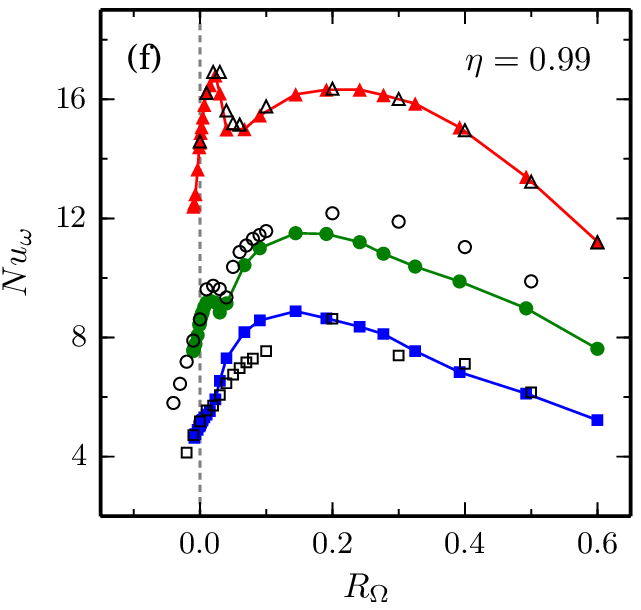}}
  \caption{Variation of the momentum transport ($Nu_\omega)$ with changing mean rotation ($R_\Omega$) calculated in DNS for the radius ratios $\eta$ specified in the sub-plots. The simulations are performed for $Re_S=5\e{3}$ ({\color{blue} $\blacksquare$}), $Re_S=10^{4}$ ({\color{OliveGreen}\Large \textbullet}) and $Re_S=2\e{4}$ ({\color{red} \large $\blacktriangle$}). The open symbols in (f) show results from simulations of RPCF. Note the increased $R_\Omega$-scale in sub-figures (a) and (b).}
\label{fig:Nu-Res}
\end{figure}

We now turn to the rotation dependence of the mean torque with increasing $\Rey_S$ and with variations in radius ratio. Figure \ref{fig:Nu-Res} shows the Nusselt number $Nu_\omega$, the torque in units of its laminar value. 
For the wide-gap TC system with $\eta=0.5$, the torque maximum initially occurs at $R_\Omega\sim0.3$ for $\Rey_S=5\e{3}$ and then shifts to $R_\Omega=0.26$ for $\Rey_S=2\e{4}$ corresponding to $\mu=-0.20$ as analysed before \citep{Brauckmann2013a,Merbold2013}. 
For $\eta=0.71$ this shift in the maximum is more pronounced, with $R_\Omega$ varying from $0.27$ at $\Rey_S=5\e{3}$ to $0.11$ at $\Rey_S=2\e{4}$ (figure \ref{fig:Nu-Res}b).
For even higher $\Rey_S$, the experiments of \citet{Paoletti2011} and \citet{VanGils2011,VanGils2012} show no further shift in the position of the torque maximum so that $R_\Omega=0.11$ is likely to be close to the asymptotic value for high $\Rey_S$.
As discussed in \S \ref{sec:burst}, these high-$\Rey_S$ torque maxima for $\eta=0.5$ and $0.71$ coincide with the onset of intermittent bursts in the outer partition.

The situation changes drastically for low-curvature TCF with $\eta\geq 0.9$ as shown in figures \ref{fig:Nu-Res}(d)--(f): instead of a single maximum, one notes 
a broad torque maximum near $R_\Omega=0.2$ and a second narrow maximum near $R_\Omega=0.02$. This narrow maximum emerges with increasing shear and, at $\Rey_S=2\e{4}$, is similar in magnitude to the broad maximum.
Indications for the narrow maximum were first seen by \citet{Ravelet2010}\footnote{The rotation number $\mbox{\textit{Ro}}$ employed by these authors is of opposite sign, i.e. $\mbox{\textit{Ro}}=-R_\Omega$.} 
in a low-curvature TC experiment ($\eta=0.917$): Their data show a very slight bump in the torque near $R_\Omega=0.02$ for $\Rey_S=1.4\e{4}$ and $1.7\e{4}$ and a monotonic increase with $R_\Omega$ for $R_\Omega\gtrsim0.05$. However, their figure 8 only shows torque measurements for $R_\Omega<0.125$ so that the broad maximum found here at $R_\Omega=0.2$ lies outside their investigated range.
The presence of two maxima, a broad one and a narrow one that emerges with increasing $\Rey_S$, was also observed in RPCF \citep{Salewski2015}. They compare well to the $Nu_\omega$ maxima in low-curvature TCF as demonstrated for $\eta=0.99$ in figure \ref{fig:Nu-Res}(f).

\begin{figure}
  \centerline{\includegraphics{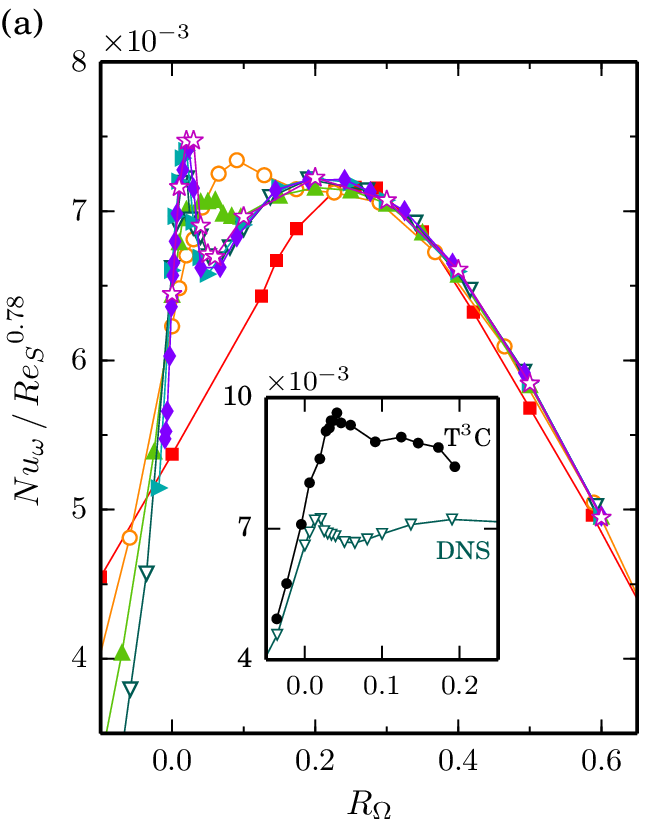}\includegraphics{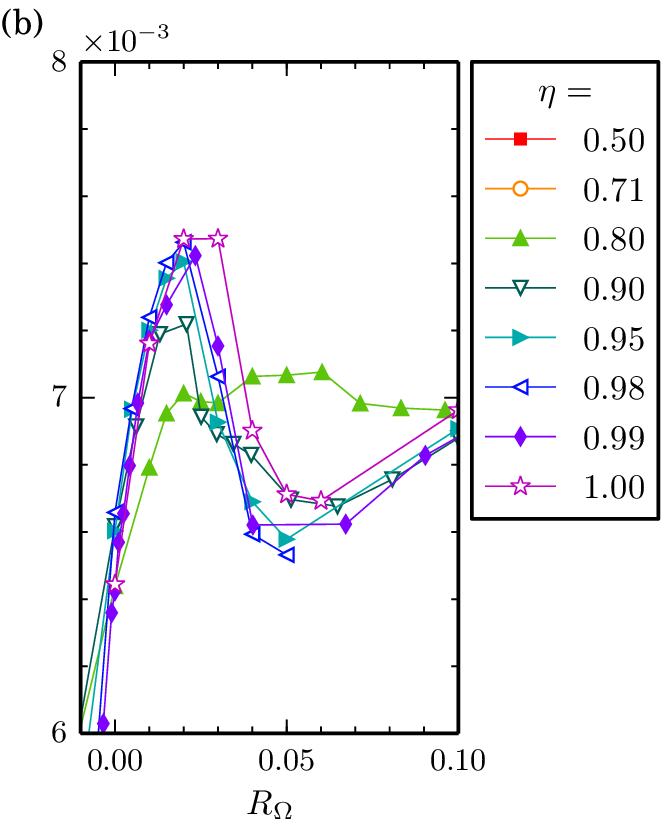}}
  \caption{Comparison of the momentum transport ($Nu_\omega$) for various radius ratios as a function of the mean rotation ($R_\Omega$) at $\Rey_S=2\e{4}$. The values are compensated with a scaling in $Re_S$ employing the scaling exponent measured by \citet{VanGils2012}. Sub-figure (b) details the narrow maximum in (a). For purposes of clarity, the results for $\eta=0.98$ are only included in (b) where the ones for $\eta=0.5$ and $\eta=0.71$ are left out. The open stars labelled by $\eta=1.00$ show results from RPCF simulations. 
  The inset in (a) compares our DNS results for $\eta=0.9$ to the T$^3$C-experiment with $\eta=0.909$ and $\Rey_S$ of a few $10^5$ \citep{Ostilla2014} and reveals a slight shift of the maximum with increasing $\Rey_S$.}
\label{fig:Nu-Ro}
\end{figure}

Finally, the torque for $\eta=0.8$ (figure \ref{fig:Nu-Res}c) shows a behaviour intermediate between that of the 
narrow-gap and the wide-gap TC systems. While the broad torque maximum known from systems with larger radius ratios is present, the second maximum around $R_\Omega=0.02$ is not as narrow and as clearly visible as for $\eta\geq0.9$. 
On the other hand, the extrapolation of equation \eqref{eq:predict}, which captures the maximum for $\eta=0.5$ and $0.71$, to the case $\eta=0.8$ predicts a maximum at $\mu_p=-0.419$ and $R_{\Omega,p}=0.069$. 
This is close to the low-curvature narrow maximum and suggests that the two will interfere.
Indeed, the magnification in figure \ref{fig:Nu-Ro}(b) of the region in $R_\Omega$ around the low-curvature narrow peak shows both, a slight maximum at $R_\Omega=0.02$ and a broader one around $R_\Omega=0.05$. 
 
In figure \ref{fig:Nu-Ro}, we show the torque as a function of the rotation number for various radius ratios.
In order to compensate for the slightly different $\Rey_S$ of the compared torques, we use the observation by \citet{Dubrulle2005} that the torque variation can be decomposed into a scaling with $\Rey_S$ modulated by an amplitude that describes the dependence on $R_\Omega$ and we thus normalise 
the torques by ${\Rey_S}^{0.78}$.
For $\eta\geq0.9$, the Nusselt numbers for different radius ratios collapse.
The figure also contains data for RPCF which also agree nicely with those of low-curvature TCF.
Data collapse is also observed with other quantities.
\citet{Dubrulle2005} previously noted that the torque $G$ normalized by its value for stationary outer cylinder $G(\mu=0)$ varies only little with $\eta$ when plotted as function of $R_\Omega$. Similarly, \citet{Paoletti2012} found a collapse of another torque ratio $G/G(R_\Omega=0)$ as a function of $R_\Omega$ for various radius ratios. 
Our figure \ref{fig:Nu-Ro} shows that up to a scaling with $\Rey_S$ a normalisation by the laminar torque suffices to compensate for the curvature dependence and to achieve the collapse for $\eta\geq 0.9$. 
Note that $\mu=0$ corresponds to $R_\Omega=1-\eta$ so that the normalisation by $G(\mu=0)$ has an additional $\eta$-dependence compared to the normalisation by $G(R_\Omega=0)$. The collapse of the data in all three cases is of similar quality and does not favour one particular normalisation.
However, plots with respect to $\mu$ (not shown here) do not give such a collapse,
confirming that $R_\Omega$ is the appropriate parameter in which to describe the transition from TCF to RPCF.

For lower $\eta$, the collapse of the torques is limited to the region $R_\Omega\gtrsim0.1$ for $\eta=0.8$ and to $R_\Omega\gtrsim0.25$ for $\eta=0.5$ and $0.71$. For these rotation numbers, turbulent bursting does not occur in the outer partition.
For smaller $R_\Omega$, where the bursting in the outer partition occurs, the Nusselt numbers in figure \ref{fig:Nu-Ro}(a) depend on $\eta$, suggesting that only this radial flow partitioning introduces the strong curvature dependence of the torques. 
In conclusion, the impact of the global rotation (parametrised by $R_\Omega$) on the turbulent momentum transport becomes independent of $\eta$ if the flow is not partly stabilised by counter-rotating the outer cylinder or if this curvature effect becomes negligible for large radius ratios.

\begin{figure}
  \centerline{\includegraphics{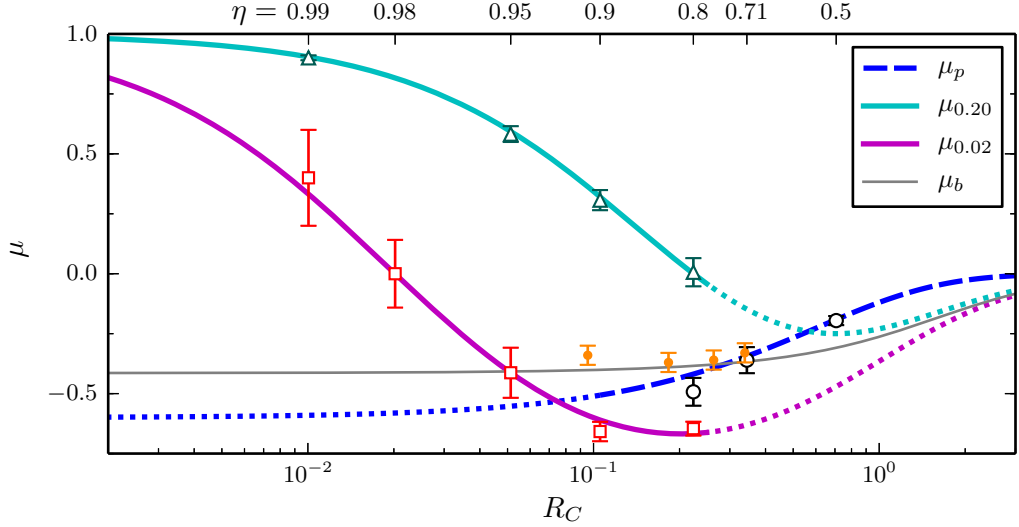}}
  \caption{Ratio of angular velocities of maximal torque $\mu_\mathrm{max}$ as a function of the curvature number $R_C=(1-\eta)/\sqrt{\eta}$. The open circles, triangles and squares summarise our numerical results for the detachment, broad and narrow maximum, respectively, at $\Rey_S=2\e{4}$ from figure \ref{fig:Nu-Ro}. 
  The maximum locations $\mu_\mathrm{max}$ were calculated from quadratic fits to the peaks in $Nu_\omega(\mu)$, except for the narrow maximum for $\eta\geq0.95$, which is not well approximated by a parabola and therefore determined by the location of the largest data point.
  The small circles indicate the experimental results by \citet{Ostilla2014} for $\Rey_S\sim10^5$ to $10^6$. These are complemented by the prediction $\mu_p(\eta)$ from equation \eqref{eq:predict} \citep{Brauckmann2013a} and by the angle bisector $\mu_b(\eta)$ by \citet{VanGils2012}. In addition, two $\mu(\eta)$-lines for constant rotation number $R_\Omega=0.2$ and $R_\Omega=0.02$ are shown, corresponding to the broad and narrow torque maxima.}
\label{fig:max-pred}
\end{figure}

The position of the torque maxima  for $\Rey_S=2\e{4}$ and $\eta\geq 0.9$, which in $R_\Omega$ is given by constant rotation numbers $R_\Omega=0.2$ and $R_\Omega=0.02$, translates into $\eta$-dependent rotation ratios $\mu$, which are represented by the thick solid lines in figure \ref{fig:max-pred}. 
The torque maximum locations $\mu_\mathrm{max}$ from the DNS (triangles and squares) closely follow these lines for $\eta\geq 0.9$. 
In particular, this reveals that for $\eta=0.99$ both the broad and the narrow torque maximum occur for co-rotating cylinders, i.e. $\mu>0$. Therefore, the narrow maximum cannot be related to the detachment mechanism that causes the torque maximum at counter-rotation for $\eta=0.5$ and $0.71$. 
For the latter radius ratios, the maximum location $\mu_\mathrm{max}$ agrees with the predictive line $\mu_p(\eta)$ in figure \ref{fig:max-pred}. 
Finally, the intermediate radius ratio $\eta=0.8$ shows indications of all three torque maxima close to the lines in figure \ref{fig:max-pred}, with the detachment and the narrow maximum lying close together and being not as pronounced as the broad maximum, cf. figure \ref{fig:Nu-Ro}.

In addition, figure \ref{fig:max-pred} shows the torque maxima identified for various radius ratios in experiments at much higher Reynolds numbers \citep{Ostilla2014}. While the maximum positions $\mu_\mathrm{max}$ for $\eta=0.714$, $0.769$ and $0.833$ conform with the trend of the detachment maximum, the one for $\eta=0.909$ clearly deviates from both $\mu_p$ and our simulations; this will be discussed in detail at the end of this section. 
Moreover, the angle bisector line $\mu_b(\eta)$, that was suggested by \citet{VanGils2012} for the position of the maximum $\mu_\mathrm{max}$, lies close to the experimental torque maxima. However, $\mu_b(\eta)$ clearly deviates from the torque maximum for $\eta=0.5$ \citep{Merbold2013} and differs in its functional behaviour for $\eta\rightarrow1$ from our DNS results.

For the range of shear Reynolds numbers  $\Rey_S\sim10^4$ investigated here, the rotation dependence of the torque significantly changes with increasing $\Rey_S$, c.f. figure \ref{fig:Nu-Res}. It is likely that this transformation process continues beyond $\Rey_S=2\e{4}$. 
However, concerning the detachment torque maximum for $\eta=0.5$ and $0.71$ it was shown that, after an initial transformation with increasing shear, the rotation dependence of the simulated torques at $\Rey_S=2\e{4}$ already compares well to experimental torque measurements at much higher $\Rey_S\sim 10^5$ to $10^6$ \citep{Brauckmann2013,Merbold2013}.
On the other hand, concerning the two torque maxima in low-curvature TCF ($\eta\gtrsim0.9$), \citet{Ostilla2014} find in their torque measurements for $\eta=0.833$ and $0.909$ that the power-law exponent $\gamma$ of the torque scaling $Nu_\omega\sim{\Rey_S}^\gamma$ depends on the system rotation for $\Rey_S$ below a few $10^5$, which indicates that the transformation of the rotation dependence still continues up to $\Rey_S\sim10^5$. 
As regards the second narrow torque maximum, we expect that it grows further in relation to the broad maximum for $\Rey_S>2\e{4}$ and that this may shift the torque-maximising rotation number a bit. 

The torque maximum from the experiment at $\eta=0.909$ \citep{Ostilla2014} supports this assumption as shown in the inset of figure \ref{fig:Nu-Ro}(a) which compares experiment and DNS: The experimental torque-maximising rotation ratio $\mu_\mathrm{max}=-0.34$ corresponds to $R_\Omega=0.04$ which only slightly deviates from $R_\Omega=0.02$ for the narrow maximum at lower $\Rey_S=2\e{4}$. In conclusion, apart from small changes with $\Rey_S$, our simulations already capture the beginning of the general turbulent behaviour at much higher shear rates.

\subsection{Torque due to vortical motion}
In this and the following sections, we investigate the momentum transport characteristics that underlie the rotation dependence of the Nusselt number discussed above. 
Vortical motions, such as Taylor vortices and their turbulent remnants, are known to effectively contribute to this transport by moving fast fluid from the inner cylinder outwards and slow outer fluid inwards. 
Their effects on the turbulence have been discussed previously \citep{Lathrop1992,Lewis1999,Martinez-Arias2014}. 
\citet{Brauckmann2013a} quantified the link of the vortical motion to the torque and found that the torque contribution of the mean vortical motion was largest near the torque maximum. Flow visualizations of DNS by \citet{Salewski2015} show distinct vortical states which underlie the narrow and broad maxima for RPCF. 
Furthermore, using experiments at $\Rey_S\sim 10^{6}$ \citep{Huisman2014} and DNS at $\Rey_S\sim 10^{5}$ \citep{Ostilla-Monico2014b}, these structures have been detected over a range of rotation numbers where their presence is associated with a single torque maximum.

To investigate the effects of vortical structures to the torque, we first extract the mean vortical motion underlying the turbulence in the full simulation and measure its contribution to the momentum transport. The mean vortices consist of temporally and streamwise averaged turbulent Taylor vortices and are analogous to the ``large-scale circulation'' found in turbulent Rayleigh-B{\'e}nard convection \citep{Ahlers2009}.
Next, since Taylor vortices do not depend on the streamwise direction, we also consider the torque which results from streamwise independent flow. We then summarize our results in figure~\ref{fig:Nu-compare}.

\subsubsection{Mean vortical motion}
\label{sec:LSC}
We follow the procedure introduced in \citet{Brauckmann2013a} and decompose the flow of the full DNS into a mean contribution \mbox{$\overline{\bv{u}}=\avg{\bv{u}}_{\varphi,t}$} (or \mbox{$\overline{\bv{u}}=\avg{\bv{u}}_{x,t}$} for RPCF) that includes the mean vortical motion in the $rz$-plane ($yz$-plane) and into the turbulent fluctuations around these mean vortices $\bv{u}''=\bv{u}-\overline{\bv{u}}$.
Substituting this velocity decomposition into the momentum flux equation \eqref{eq:contTC} results in a separation of the torque into a contribution that is caused by the mean vortical motion $\overline{G}$ and a second contribution that is due to turbulent fluctuations $G''$, i.e. $G=\overline{G}+G''$ with
\begin{eqnarray}
 \overline{G} &=& \frac{\Rey_S^{2}}{r_2-r_1} \int_{r_1}^{r_2}
 \left(\left<\overline{u}_r\overline{\mathcal{L}}\right>_{\varphi z,t}
 -\Rey_S^{-1} r^2 \partial_r\left<\overline{\omega}\right>_{\varphi z,t} \right)  \, r\,\mathrm{d}r , \nonumber \\
 G'' &=& \frac{\Rey_S^{2}}{r_2-r_1} \int_{r_1}^{r_2} \left<u_r''\mathcal{L}''\right>_{\varphi z,t} \, r\,\mathrm{d}r .
 \label{eq:G_contrib}
\end{eqnarray}
Analogous expressions for the decomposition of the driving force into $F=\overline{F}+F''$ can be obtained for RPCF by substituting the velocity decomposition into the momentum flux equation \eqref{eq:contPC}. To capture the correct amplitude of the turbulent Taylor vortices, we fix their spanwise position during the temporal average of $\overline{\bv{u}}$ and in \eqref{eq:G_contrib} by correcting a potential spanwise drift. Such a drift would lead to a cancellation of the large-scale motion over time, even though it might be strong.
Note that while the total transport $G$ is constant over $r$, the individual terms in \eqref{eq:G_contrib} vary with the radius. This motivates the additional radial average in \eqref{eq:G_contrib} in order to quantify the typical strength of the torque contributions. Since we are interested in the mean vortical motion which dominates in the central region and not in the boundary layers, we restrict this average to the range between $(r_1-r_i)/d=1/4$ and $(r_2-r_i)/d=3/4$. 

In figure \ref{fig:Nu-compare}, the torque due to the large-scale circulation $\overline{G}$ is compared to the total torque $G$. For all cases, $\overline{G}$ exhibits a maximum. 
For $\eta\geq0.8$ and RPCF, the maximum of $\overline{G}$, where the momentum transport by the mean vortices is most effective, nearly coincides with the narrow maximum of the total torque $G$. Moreover, it is apparent that a distinct narrow peak also occurs in mean-vortex torque $\overline{G}$ for large $\eta$. This suggests that the narrow, high-$\Rey_S$ maximum is linked to an efficient large-scale circulation.
For lower radius ratios, i.e. $\eta=0.5$ and $0.71$, the coincidence between the maxima in $\overline{G}$ and $G$ is not as significant; nevertheless, as the analysis in \cite{Brauckmann2013a} demonstrates, this coincidence does exist for $\eta=0.5$ and $0.71$ when the radial average in \eqref{eq:G_contrib} covers the complete radial gap instead of the central region. 
Finally, the mean-vortex torques $\overline{G}$ in figure \ref{fig:Nu-compare} reveal that the mean Taylor vortices occur and grow in strength for $R_\Omega>0$ as previously observed in the experiment by \citet{Ravelet2010}.

\subsubsection{Importance of streamwise invariant structures}
\begin{figure}
  \centerline{\includegraphics{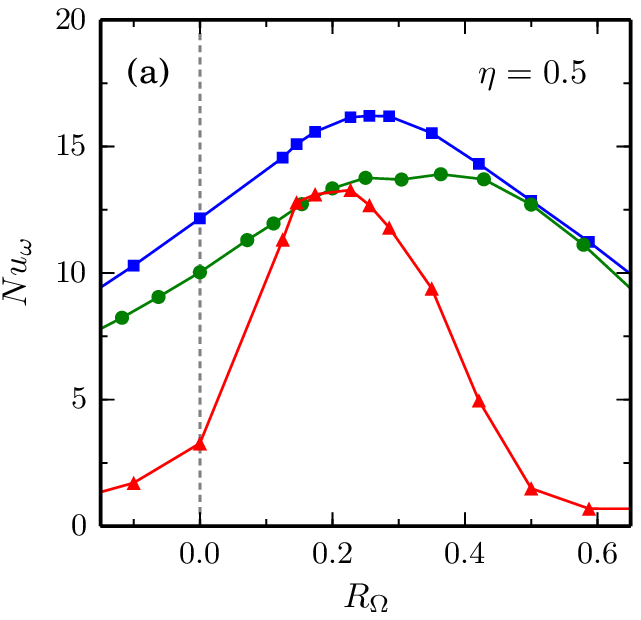}\includegraphics{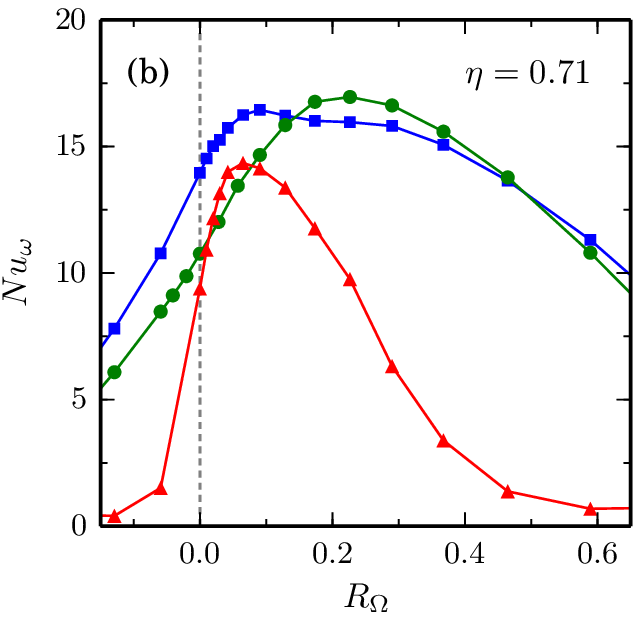}}
  \centerline{\includegraphics{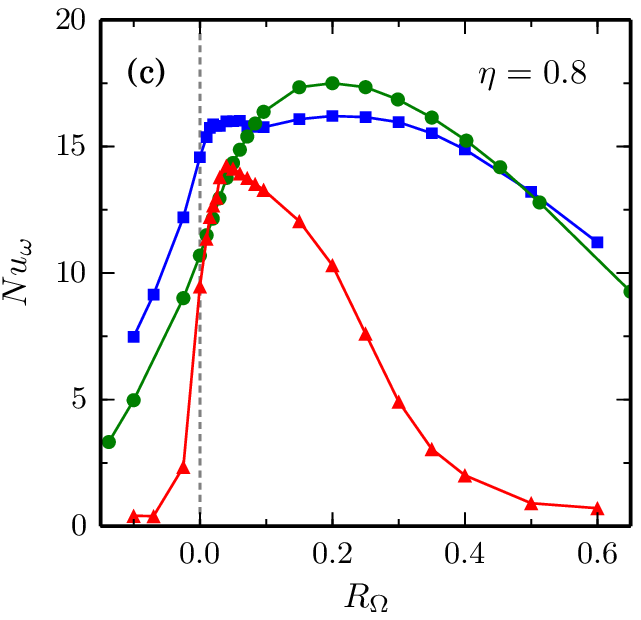}\includegraphics{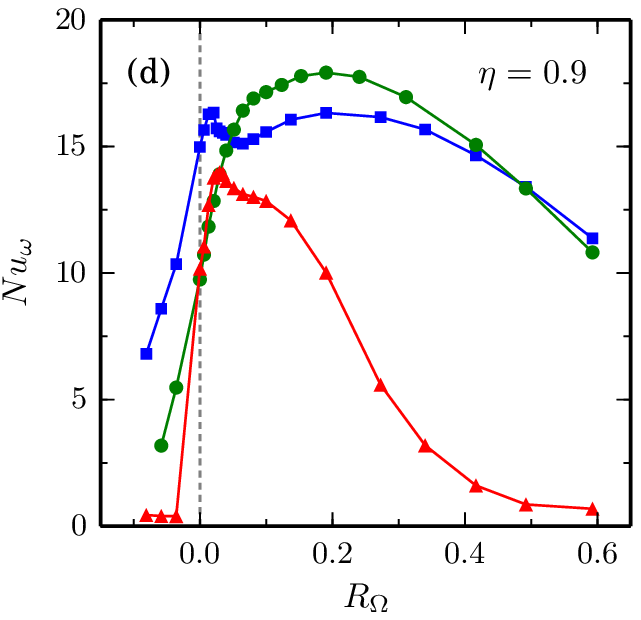}}
  \centerline{\includegraphics{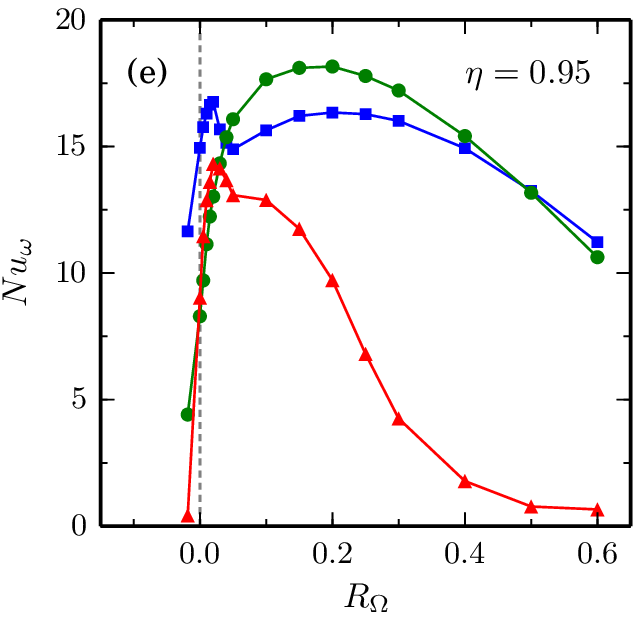}\includegraphics{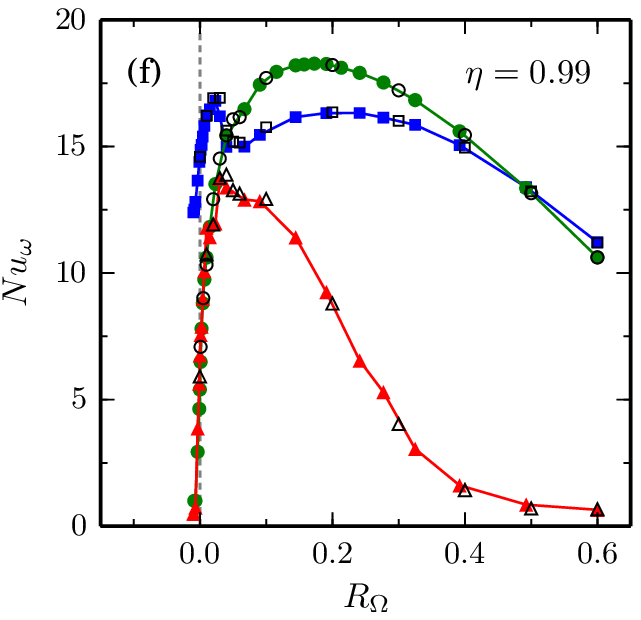}}
  \caption{Comparison of the total torque $G$ ({\color{blue} $\blacksquare$}) to the torque that results from the streamwise invariant flow $G_\mathrm{2D}$ ({\color{OliveGreen}\Large \textbullet}) and to the torque due to the mean flow $\overline{G}$ ({\color{red} \large $\blacktriangle$}) for $\Rey_S=2.0\e{4}$. All torques are measured in units of the laminar torque $G_\mathrm{lam}$ giving Nusselt numbers $Nu_\omega$. 
  The open symbols in (f) show corresponding results from RPCF simulations.}
\label{fig:Nu-compare}
\end{figure}

We perform simulations that force the flow to be invariant in the downstream direction by taking no Fourier mode in the $\varphi-$direction ($x-$direction for RPCF). 
For TCF this corresponds to axisymmetric simulations. 
Note that this procedure differs from $2D$ simulations in a cross-section because three velocity components are still active, however, allowing only for spatial variations in two directions (wall-normal and spanwise). These simulations of streamwise invariant vortical flow result in torques, denoted as $G_\mathrm{2D}$, that we compare to the total torques $G$ in figure \ref{fig:Nu-compare}.
The torque $G_{\mathrm{2D}}$ also shows a broad maximum, which has an $\eta$-dependent shift in its location: for $\eta \sim 1$, the maximum is at $R_\Omega = 0.18$; and for $\eta = 0.5$, it is near $R_\Omega = 0.35$.
For some radius ratios, namely $\eta = 0.8$, $\eta = 0.9$, this places the $G_{\mathrm{2D}}$-maximum at nearly the same $R_\Omega$ as the broad maximum in the total torque; however, the streamwise invariant simulations overestimate the amplitude of the maximum.
In fact, for $\eta \geq 0.8$ and RPCF, the $G_{\mathrm{2D}}$-maximum resembles the broad maximum in the total torque.
Furthermore, it also appears to agree with the plateau that occurs in the total torque beside the actual torque maximum for $\eta = 0.71$ (figure \ref{fig:Nu-compare}b).
This suggests that the plateau for $\Rey_S=2\e{4}$ as well as the broad maximum at lower $\Rey_S$ (cf. figure \ref{fig:Nu-Res}b) follow from the same mechanism that causes the broad maximum for $\eta\geq0.8$.

The analysis reveals several features of the contribution of vortical motion to the torque for different degrees of curvature. The streamwise-averaged mean flow ($\overline{G}$) appears to reproduce the narrow torque maximum, but this is not reflected by the streamwise-invariant flow ($G_\mathrm{2D}$). 
This seems to be counter-intuitive when considering that $\overline{\bv{u}}$ is streamwise invariant like the simulations underlying $G_\mathrm{2D}$, but neither the detachment maximum in the total torque for $\eta\leq0.8$ nor the narrow maximum for $\eta\geq0.8$ occur in the torques from the streamwise invariant simulations. 
As a consequence, these maxima must arise from more complicated flows which allow for streamwise fluctuations and cause an additional increase of the momentum transport.
The reasons for the strong momentum transport differ between the broad and narrow torque maximum as section \S \ref{sec:fluct} will illustrate: strong streamwise vortices as well as a high correlation between the radial flow $u_r$ and the angular momentum $\mathcal{L}$ in these vortices are required for an effective momentum transport and, thus, high torque. 

\section{Flow characterisation}
\label{sec:charact}
In section \ref{sec:Nusselt} we discussed that the momentum transport converges to a universal behaviour for the low-curvature TCF ($\eta\geq0.9$), in that the rotation dependence becomes independent of $\eta$. Furthermore, the dependence on $R_\Omega$ agrees with that observed in RPCF. 
In the following, we analyse how the convergence for $\eta\geq0.9$ extends to other flow characteristics such as mean profiles and turbulent fluctuations. Moreover, we compare our TC results to the behaviour in RPCF. 

\subsection{Angular momentum profiles}
\label{sec:prof}
\begin{figure}
  \centerline{\includegraphics{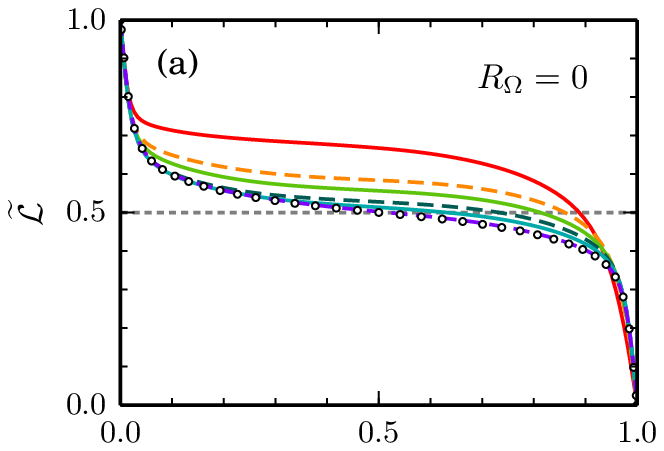}\includegraphics{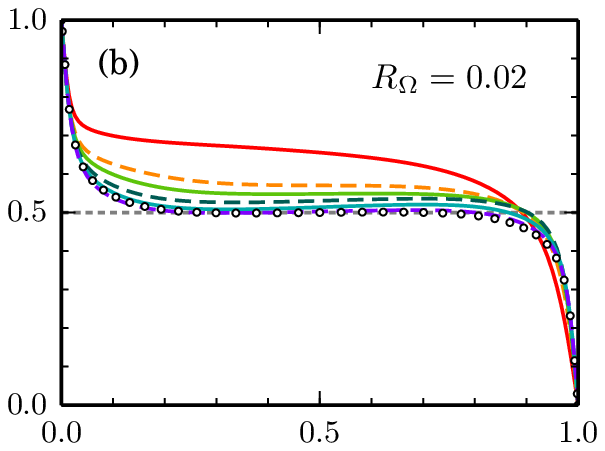}}
  \centerline{\includegraphics{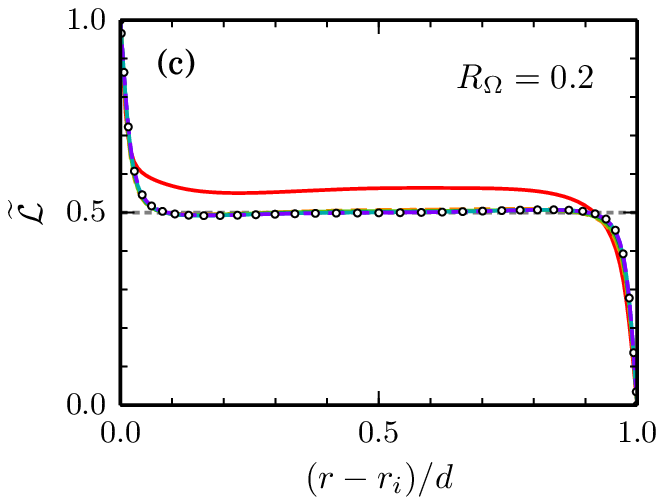}\includegraphics{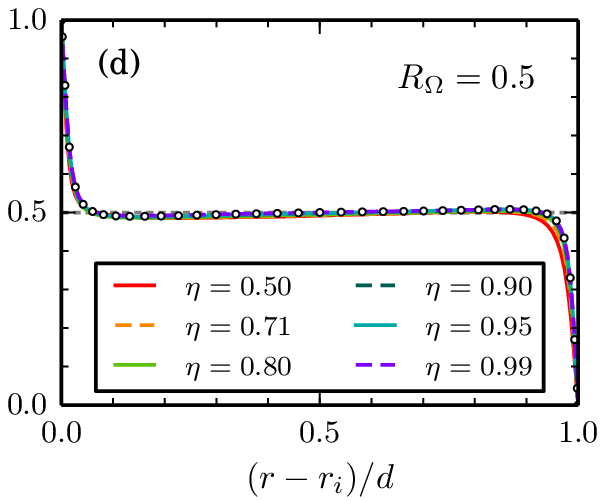}}
  \caption{Profiles of the rescaled specific angular momentum 
  $\widetilde{\mathcal{L}} = \left(\avg{\mathcal{L}} -\mathcal{L}_o\right) / \left(\mathcal{L}_i -\mathcal{L}_o\right)$ for the mean system rotations defined by $R_\Omega$ and for six different radius ratios. 
  The corresponding RPCF profile $\avg{u_x}-R_\Omega y$ rescaled to the interval $[0,1]$ is shown by the dots.
  Each figure contains profiles for $\eta=0.5$, $0.71$, $0.8$, $0.9$, $0.95$, and $0.99$ as listed in (d). 
  Some of the profiles collapse, especially in (c) and (d). In (a) and (b) the profile centre level decreases monotonically with increasing $\eta\rightarrow1$. The dotted line shows the arithmetic mean angular momentum $\widetilde{\mathcal{L}}=0.5$.}
\label{fig:prof}
\end{figure}

We analyse here the mean profiles of the specific angular momentum $\mathcal{L}=ru_\varphi$ which is radially transported between the cylinders according to the continuity equation \eqref{eq:contTC}. 
To compensate for the varying global system rotation, we show rescaled profiles $\widetilde{\mathcal{L}} = \left(\avg{\mathcal{L}} -\mathcal{L}_o\right) / \left(\mathcal{L}_i -\mathcal{L}_o\right)$ in figure \ref{fig:prof} where $\mathcal{L}_i$ and $\mathcal{L}_o$ denote the specific angular momentum of the inner and outer cylinder.
As a result, the profiles for various radius ratios $\eta$ that belong to the same rotation number $R_\Omega$ collapse as long as the flow is not subject to the radial partitioning of stability for moderate counter-rotation (cf. section \ref{sec:burst}): 
The rotation number $R_\Omega=0.5$ corresponds to co-rotation ($\mu\geq0$) for all investigated $\eta$, and all profiles of $\widetilde{\mathcal{L}}$ in figure \ref{fig:prof}(d) collapse well. 
However at $R_\Omega=0.2$, only the simulation for $\eta=0.5$ with $\mu=-0.22$ shows the bursting in the outer partition and a deviating profile in figure \ref{fig:prof}(c). 
Similarly at $R_\Omega=0.02$ (figure \ref{fig:prof}b), the profiles for $\eta\leq0.9$ corresponding to $\mu$-values in the bursting regime clearly deviate from the profiles for $\eta=0.95$ and $0.99$ with $\mu>\mu_p$ which are closer together.
Finally, no collapse is observed for $R_\Omega=0$ which always corresponds to exact counter-rotation.
In addition, figure \ref{fig:prof} shows $u_x$ profiles from RPCF with the system rotation subtracted, i.e. $\avg{u_x} - R_\Omega y$, that agree well with the collapsed $\widetilde{\mathcal{L}}$ profiles from TCF when rescaled to the same interval. The reason for this correspondence between angular momentum profiles in TCF and $\avg{u_x} - R_\Omega y$ in RPCF will be discussed in the next section.

The $\widetilde{\mathcal{L}}$ profiles tend towards a universal shape for increasing $R_\Omega$, as the radial partitioning of stability disappears. The specific angular momentum in the centre becomes nearly flat (with a slightly positive slope) and approaches $\widetilde{\mathcal{L}}=0.5$ which corresponds to $\avg{\mathcal{L}}=(\mathcal{L}_i+\mathcal{L}_o)/2$ and indicates that the angular momentum is well mixed. 
Already \citet{Wendt1933} noted that in the unstable regime the angular momentum becomes almost constant in the centre which was also repeatedly observed for the case of stationary outer cylinder \citep{Wattendorf1935,Taylor1935,Smith1982,Lewis1999,Dong2007}. Flat $\mathcal{L}$ profiles resemble the situation in Rayleigh-B\'{e}nard convection where the transport of heat results in a well mixed temperature in the bulk \citep{Tilgner1993,Kerr1996,Brown2007,Ahlers2012}.

\subsection{Mixing of angular momentum and neutral stability}
\label{sec:slope}
The observation that the specific angular momentum $\mathcal{L}=r u_\varphi$ becomes almost flat in the middle is remarkable and calls for an explanation. Moreover, it is of interest whether an analogous phenomenon can be found in RPCF. We will approach both questions by the stability properties of the mean profile around which the turbulence fluctuates.
As long as the mean profile ($\avg{\omega}(r)$ in TCF and $\avg{u_x}(y)$ in RPCF) is unstable due to a strong a gradient in the centre, it further enforces turbulence which increases the mixing of angular (or linear) momentum; the mixing flattens and therefore re-stabilises the profile. However, a stable, low-gradient profile decreases turbulent mixing, so that the profile becomes steeper and eventually unstable again. 
Therefore, we assume that both processes balance in the statistically averaged sense and the mean turbulent profile becomes neutrally stable in the centre. 

For TCF with stationary outer cylinder, \cite{Wattendorf1935} and \cite{Taylor1935} already noted that the mean angular momentum profile in the middle tends towards one of neutral stability according to Rayleigh's inviscid criterion \citep{Rayleigh1917}, which predicts a constant angular momentum profile for this case. Neglecting viscous effects in the central part of the flow becomes justifiable at high Reynolds numbers.
The flat $\widetilde{\mathcal{L}}$ profiles in figure \ref{fig:prof} indeed realise Rayleigh's neutral stability criterion in the centre, thus, supporting the aforementioned assumptions. More precisely, Rayleigh's discriminant for neutral stability \citep{Chandrasekhar1961} reads
\begin{eqnarray}
 0\equiv\frac{2\omega}{r}\,\p_r\left(r^2\omega\right) &=&
 \left(R_\Omega-1+\frac{\widetilde{r}^2}{r^2}\right)\frac{1}{r} \p_r\mathcal{L} 
    \label{eq:Lstab}\\
 &=&  \left( R_\Omega-1+\frac{\widetilde{r}^2}{r^2}\right)
      \left( R_\Omega-1+\frac{\widetilde{r}^2}{r^2} + r \p_r\omega\right)
    \label{eq:Omstab}
\end{eqnarray}
here expressed so that it serves as a condition for a neutrally stable turbulent mean profile 
in either $\mathcal{L}$ or $\omega$ in the middle. For that purpose, we substituted the angular velocities representing the mean profile by the laminar solution $\omega_\mathrm{lam}(r)$ from equation \eqref{eq:omega_lam} because in the centre $\omega_\mathrm{lam}$ provides a reasonable 
approximation to the magnitude of the turbulent mean profile as evidenced by the comparison in \citet{Brauckmann2013}. 
Since in the middle $r$ is close to the geometric radius $\widetilde{r}=\sqrt{r_i r_o}$, equation \eqref{eq:Omstab} implies that $r\partial_r\omega\approx-R_\Omega$ for $R_\Omega\neq0$. \citet{Ostilla2013} obtained a similar linear relationship by assuming that the Coriolis force term balances the convective term in the $\varphi$-component of the equations of motion \eqref{eq:NS-TC}.
Equations (\ref{eq:Lstab}) and (\ref{eq:Omstab}) then show that both force balance \citep{Ostilla2013} and inviscid stability \citep{Rayleigh1917} predict a profile with flat angular momentum.

The equivalent stability formulation in RPCF was derived from an analogy between rotating shear flows and buoyant flows \citep{Bradshaw1969,Tritton1992}: The stability expressed by a quantity similar to the Richardson number in stratified flows results in the neutral stability condition
\begin{equation}
 R_\Omega\left(R_\Omega-\p_y u_x\right)\equiv0
\label{eq:Ustab}
\end{equation}
which is approximately contained in \eqref{eq:Omstab} since $\widetilde{r}/r\approx 1$ in the middle, and $r\p_r\omega$ corresponds to $-\p_y u_x$. The difference in sign originates from the fact that the radial coordinate in TCF is antiparallel to the wall-normal coordinate in RPCF, cf. figure \ref{fig:sketch}. 
Also for RPCF we assume that the turbulent mean profile in the middle becomes neutrally stable and fulfils \eqref{eq:Ustab}, as proposed previously by \citet{Suryadi2014}. 
In conclusion, the negative profile slope $-\partial_y\avg{u_x}$ in RPCF corresponds to the angular velocity gradient $r\partial_r\avg{\omega}$ in TCF. Moreover, comparing \eqref{eq:Ustab} to \eqref{eq:Lstab} reveals that the combination \mbox{$(R_\Omega - \p_y\avg{u_x})$} in RPCF plays the role of the angular momentum gradient $r^{-1}\partial_r\avg{\mathcal{L}}$ in TCF, and we therefore compare the modified RPCF profile $\avg{u_x}-R_\Omega y$ to the angular momentum profiles in figure \ref{fig:prof}.

\begin{figure}
  \centerline{\includegraphics{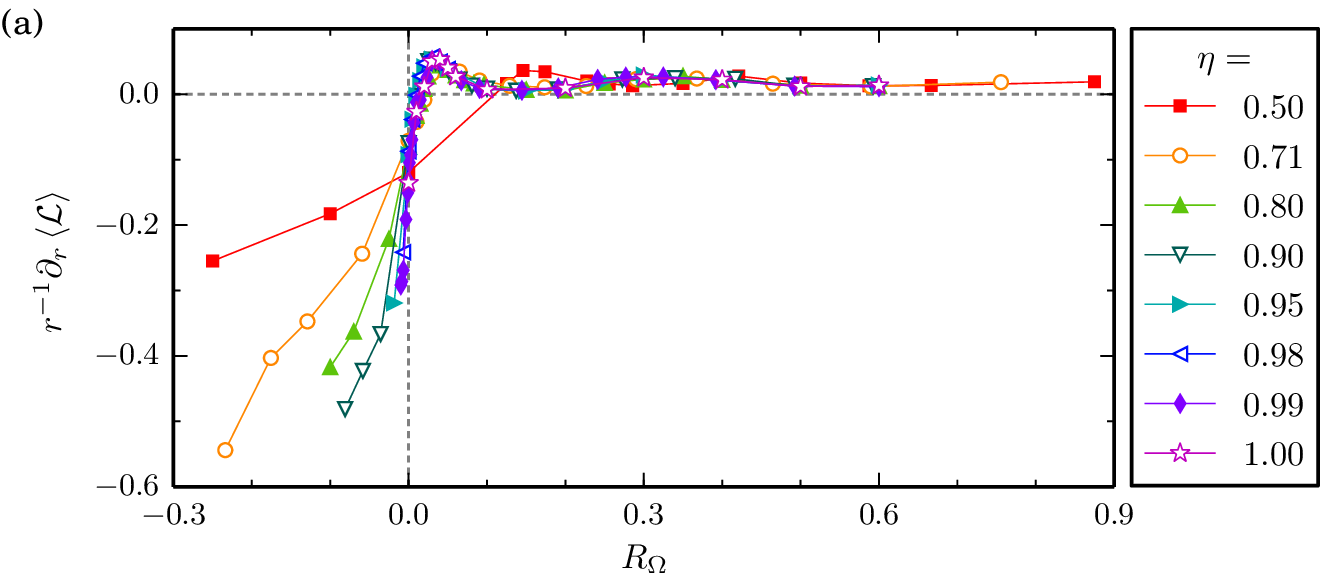}}
  \centerline{\includegraphics{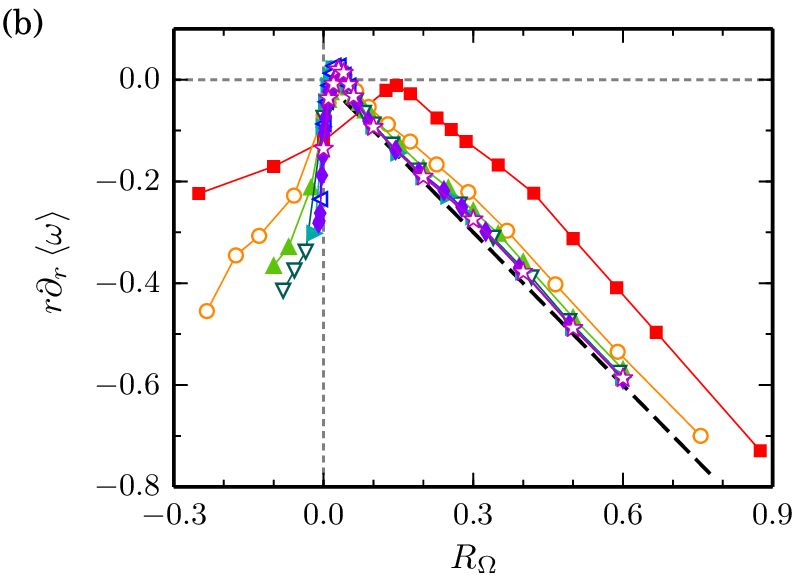}\includegraphics{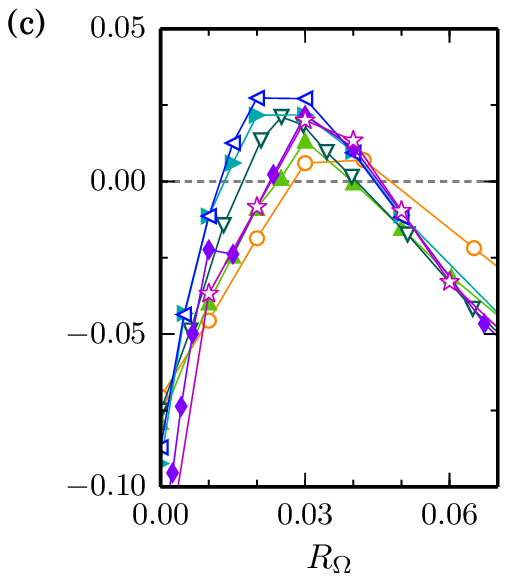}}
  \caption{Profile gradients versus the rotation number $R_\Omega$ for $Re_S=2.0\e{4}$ and various radius ratios $\eta$. (a) radial gradient of the specific angular momentum profile $\avg{\mathcal{L}}$ in TCF and the corresponding quantity $(R_\Omega-\partial_y\avg{u_x})$ for RPCF, (b) gradient of the angular velocity profile $\avg{\omega}$ in TCF and the negative downstream velocity gradient $-\p_y \avg{u_x}$ in RPCF, (c) detail of the maximum in (b). The black dashed line in (b) indicates the estimate \mbox{$r\p_r\avg{\omega}=-\p_y\avg{u_x}=-R_\Omega$} from the neutral stability conditions \eqref{eq:Omstab} and \eqref{eq:Ustab}. All gradients are averaged in the central region $(r-r_i)/d\in[0.4, 0.6]$.}
\label{fig:slope}
\end{figure}

To test the assumption of neutrally stable mean profiles, the dependence of the profile gradients on $R_\Omega$, and the proposed analogies between TCF and RPCF, we evaluate the relevant profile gradients in the central region for $\Rey_S=2\e{4}$ and various radius ratios.
Figure \ref{fig:slope}(a) shows that the specific angular momentum becomes almost flat with a slightly positive slope for $R_\Omega\gtrsim0.02$, except for $\eta=0.5$ which is well mixed only for $R_\Omega\gtrsim0.1$. 
Experiments show that the slightly positive angular momentum slope persists for much higher $\Rey_S$ \citep{Smith1982,Lewis1999}.
In addition, for most $R_\Omega>0$, the profile slopes measured by $r^{-1}\p_r\avg{\mathcal{L}}$ collapse for different $\eta$, which demonstrates the universality of the weak $R_\Omega$-dependence. Furthermore, 
the quantity $(R_\Omega-\p_y\avg{u_x})$ from RPCF also shows this high level of agreement expected from the neutral stability analogy \eqref{eq:Lstab} and \eqref{eq:Ustab}. 
In contrast, the profile gradients for $R_\Omega<0$ ($R_\Omega<0.1$ for $\eta=0.5$) clearly 
vary with the radius ratio. For these rotation numbers, a pronounced stable partition exists which introduces a strong curvature dependence as discussed for the torques in section \S \ref{sec:Nusselt}.

The flattening of the specific angular momentum implies that the angular velocity gradients depend on $R_\Omega$ as shown in figure \ref{fig:slope}(b,c) and in similar figures in \citet{Ostilla2014}. In the unstable regime with $R_\Omega>0$, the profile slopes $r\p_r\avg{\omega}$ approach the neutral stability estimate \mbox{$r\p_r\avg{\omega}=-\p_y\avg{u_x}=-R_\Omega$} as $\eta$ tends to $1$. 
Moreover for low-curvature TCF ($\eta\gtrsim0.9$), they coincide with the $u_x$-profile gradients in RPCF which supports the proposed analogy between $\avg{\omega}$ and $\avg{u_x}$. The approximate scaling $\p_y\avg{u_x}\sim R_\Omega$ in RPCF conforms with observations by \citet{Bech1997} and \citet{Suryadi2014}.
However for smaller $\eta$, the $\avg{\omega}$-gradients depend more strongly on the curvature than the $\avg{\mathcal{L}}$-gradients, consistent with the additional radial dependence in condition \eqref{eq:Omstab}. 
Furthermore, we note that a slight local super-rotation (corresponding to a shear-inversion in RPCF) occurs in the central region for $R_\Omega\approx0.03$ as detailed in figure \ref{fig:slope}(c). This inversion may result from an efficient momentum transport by the large-scale circulation moving fast fluid from the inner wall outwards and slow fluid inwards. The maximum in the mean vortex torque $\overline{G}$ around $R_\Omega\approx0.03$ indeed supports this picture, cf. figure \ref{fig:Nu-compare}.

For the low-curvature TCF with $\eta\gtrsim0.9$, the flattest $\omega$-profile (i.e. maximum in $r\p_r\avg{\omega}$) and the narrow torque maximum occur for almost the same rotation number ($R_\Omega\approx0.03$ versus $0.02$) which extends an observation in \citet{Ostilla2014} for $\eta\leq0.909$ to larger radius ratios and to the limiting case of RPCF, where the flattest $u_x$-profile is linked to the narrow maximum in $Nu_u$. 
This connection also exists for $\eta=0.8$, thus, supporting the independence of the narrow maximum at $R_\Omega=0.02$ from the detachment maximum at $R_\Omega=0.05$.
For $\eta=0.5$ and $\eta=0.71$ the $R_\Omega$-value of the flattest angular velocity profile does not coincide with the location of the torque maximum, which is another evidence that this detachment maximum differs from the narrow torque maximum observed for $\eta\gtrsim0.9$.
In contrast, \citet{Ostilla2014} found the flattest $\omega$-profile for $\eta=0.5$ at $R_\Omega=0.25$ and for $\eta=0.714$ at $R_\Omega=0.12$ which agree with the corresponding torque-maximizing rotation numbers. However, the difference might result from the fact that their values are extrapolated from simulations at $\Rey_S\lesssim5\e{3}$ lower than the $\Rey_S=2\e{4}$ used here.

The gradients of the $\omega$-profile ($u_x$-profile in RPCF) are relevant when analysing the momentum transport in the centre and the torque (force) needed to drive the walls. They measure the diffusive part of the current $J_\mathcal{L}$ ($J^u$, respectively) which is mediated by viscosity, i.e. the last term in equations \eqref{eq:contTC} and \eqref{eq:contPC}. 
A flat profile means that the convective current $J_{\mathcal{L},c}=\avg{u_r\mathcal{L}}$ ($J_{x,c}=\avg{u_y u_x}$ in RPCF) accomplishes all momentum transport in the middle.

\subsection{Turbulent fluctuations}
\label{sec:fluct}

\begin{figure}
  \centerline{\includegraphics{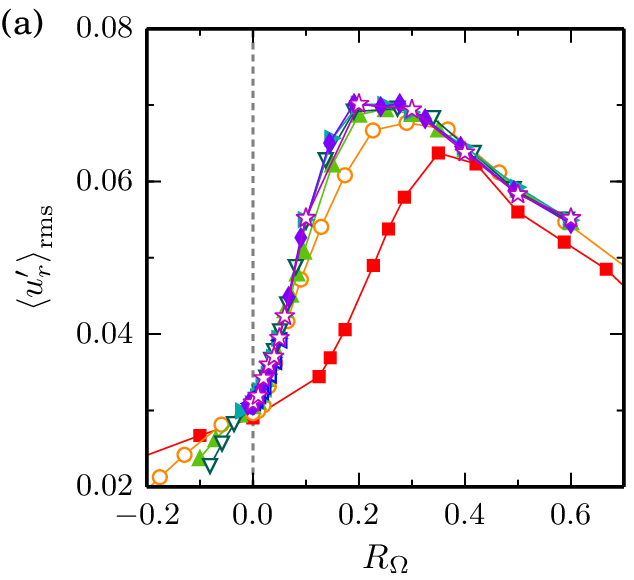}\includegraphics{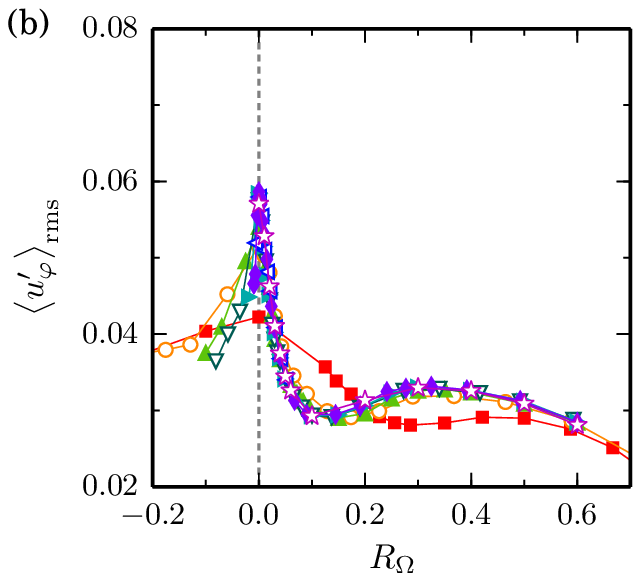}}
  \centerline{\includegraphics{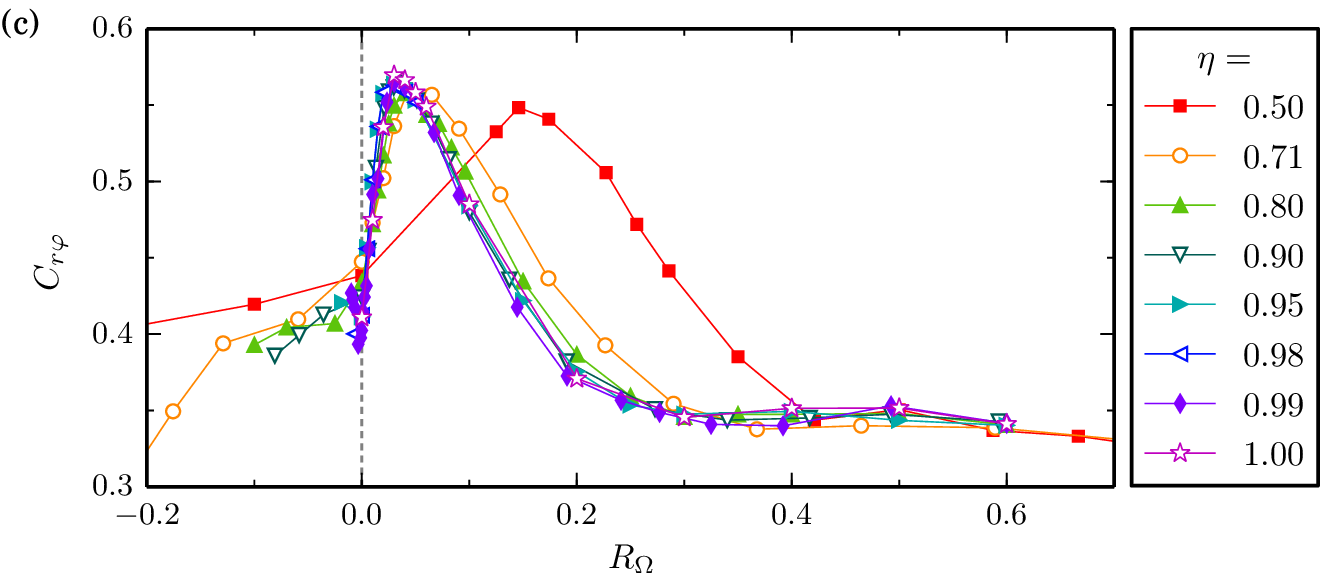}}
  \caption{Fluctuation amplitude of the wall-normal velocity $\avg{u_r'}_\mathrm{rms}$  ($\avg{u_y'}_\mathrm{rms}$ for RPCF) in (a) and of the streamwise velocity $\avg{u_\varphi'}_\mathrm{rms}$ ($\avg{u_x'}_\mathrm{rms}$) in (b) and their cross-correlation coefficient $C_{r\varphi}$ ($C_{yx}$) in (c) as a function of the rotation number $R_\Omega$ for $Re_S=2.0\e{4}$ and various radius ratios $\eta$. All quantities are radially average in the central region $(r-r_i)/d \in \left[1/3, 2/3\right]$ where momentum is mainly transported by the convective current $J_{\mathcal{L},c}=r\avg{u_r' u_\varphi'}$ ($J_{x,c}=\avg{u_y' u_x'}$).}
\label{fig:fluct}
\end{figure}

Finally, we study the effect of the mean system rotation ($R_\Omega$) on the characteristics of the convective momentum transport by turbulent fluctuations. We demonstrate that not all deviations from the laminar state contribute to the net convective momentum transport but only the turbulent fluctuations 
around the mean profile. 
This becomes apparent when introducing the velocity deviation from the mean profile
$\bv{u}'=\bv{u}-\avg{\bv{u}}$ by using the original average $\avg{\cdots} = \avg{\cdots}_{\varphi z,t}$; note that this differs from $\bv{u}''=\bv{u}-\avg{\bv{u}}_{\varphi,t}$, which was introduced in \S \ref{sec:LSC} to describe turbulent fluctuations around the mean vortical motion, see also \citet{Bilson2007}. 
As a result of incompressibility \eqref{eq:cont} the mean radial flow  vanishes, i.e. $\avg{u_r}=0$, so that $u_r'=u_r$. Moreover, by substituting the azimuthal velocity decomposition $u_\varphi=\avg{u_\varphi}+u_\varphi'$ into the convective angular momentum current,
\begin{equation}
 J_{\mathcal{L},c}=r\avg{u_r\,u_\varphi}=r\avg{u_r}\avg{u_\varphi}+r\avg{u_r\,u_\varphi'}=r\avg{u_r'\,u_\varphi'} ,
\end{equation}
we see that only the correlated fluctuations $u_r'$ and $u_\varphi'$ contribute to the net convective transport. Inspired by the analysis of \citet{Burin2010}, we measure their amplitude by area-time averaged 
root-mean-square values 
$\avg{u_r'}_\mathrm{rms}$ and $\avg{u_\varphi'}_\mathrm{rms}$ that enable the definition of a cross-correlation coefficient 
\begin{equation}
 C_{r\varphi}= \frac{\avg{u_r'\, u_\varphi'}}{\avg{u_r'}_\mathrm{rms} \avg{u_\varphi'}_\mathrm{rms}} 
 \qquad \mbox{so that} \qquad 
 J_{\mathcal{L},c}= r \avg{u_r'}_\mathrm{rms} \avg{u_\varphi'}_\mathrm{rms} C_{r\varphi}.
\end{equation}
The corresponding quantities in RPCF are
\begin{equation}
  C_{yx}= \frac{\avg{u_y'\, u_x'}}{\avg{u_y'}_\mathrm{rms} \avg{u_x'}_\mathrm{rms}} 
 \qquad \mbox{so that} \qquad 
 J_{x,c}= \avg{u_y'}_\mathrm{rms} \avg{u_x'}_\mathrm{rms} C_{yx}.
\end{equation}
Consequently, we may consider increases in the convective (angular) momentum transport as being due to violent fluctuations of the wall-normal or streamwise velocities, or due to a strong cross-correlation between them. All three effects are analysed as functions of $R_\Omega$ for $\Rey_S=2.0\e{4}$ in figure \ref{fig:fluct}.

Generally, we observe for $\eta\geq0.8$ and RPCF that the fluctuation amplitudes and correlation coefficients collapse well in the unstable range with $R_\Omega\gtrsim0$, which again indicates that low-curvature TCF and RPCF show the same rotation dependence.
The wall-normal flow grows in amplitude when rotation sets in ($R_\Omega>0$) and is suppressed by strong rotation forming a maximum around $R_\Omega=0.25$ for $\eta\geq0.71$; this nearly coincides with the broad torque maximum at $R_\Omega=0.2$ for $\eta\geq0.8$, and it resembles the maximum observed in the torque $G_\mathrm{2D}$ from streamwise invariant simulations (cf. figure \ref{fig:Nu-compare}). 
The co-occurrence of the maximum wall-normal flow with the broad peak supports the picture that the broad torque maximum is associated with strong vortical motion. 
While the co-occurrence holds for the plateau next to the actual maximum for $\eta=0.71$ (figure \ref{fig:Nu-compare}b), the correlation between the maxima is less clear for $\eta=0.5$.

On the other hand, the streamwise velocity fluctuations are strongest without system rotation, forming a sharp maximum around $R_\Omega=0$, except for the high-curvature TCF with $\eta=0.5$ (figure \ref{fig:fluct}b). Even though, a corresponding peak in the torques is missing since the narrow torque maximum at $R_\Omega=0.02$ clearly drops off towards $R_\Omega=0$ (cf. figure \ref{fig:Nu-Ro}b).

However, for $\eta\geq0.9$ the maximum in the cross-correlation coefficients at $R_\Omega\approx0.03$ (figure \ref{fig:fluct}c) comes closer to the narrow torque maximum suggesting that the latter is associated to a very efficient (angular) momentum transport by moderately strong but correlated vortices. 
Moreover, the peak in the cross-correlation coefficients coincides with the maximum in the torque $\overline{G}$ due to the streamwise-averaged mean flow (cf. figure \ref{fig:Nu-compare}). The maximum in $C_{r\varphi}$ at $R_\Omega=0.15$ and at $R_\Omega=0.05$ for $\eta=0.5$ and $\eta=0.71$, respectively, does not coincide with the torque maximum for the corresponding radius ratio indicating that the latter differs from the narrow maximum for $\eta\gtrsim0.9$.

\section{Summary and conclusions}
By describing TCF in the rotating reference-frame as proposed by \citet{Dubrulle2005}, we were able to study its limit to RPCF as the radius ratio $\eta$ approaches $1$. 
In this framework, shear and mean system rotation are quantified by the same parameters $\Rey_S$ and $R_\Omega$ in both systems and curvature effects measured by $R_C$ disappear with $\eta\rightarrow1$. 
The data collapse observed for $\eta\geq0.9$ shows that $\Rey_S$ and $R_\Omega$ are the appropriate parameters in which to describe the transition from TCF to RPCF.
Moreover, we analysed the analogy between the angular momentum transport in TCF and the linear momentum transport in RPCF by utilising the correspondence between the currents $J_\mathcal{L}$ and $J^u$ which are connected to the torque $G$ and the force $F$ in the respective geometries. 

As a consequence of curvature, the TCF for counter-rotating cylinders features a radial partitioning of stability that results in enhanced intermittent fluctuations in the outer partition. For a constant shear $\Rey_S$ and changing system rotation, the occurrence of the intermittency has been linked to the torque maximum that forms as a function of $R_\Omega$ for $\eta=0.5$ and $0.71$ \citep{VanGils2012,Brauckmann2013a}. 
Our simulations show that the onset of enhanced outer fluctuations agrees in the range $0.5\leq\eta\leq0.9$ with the prediction $\mu_p$ for the occurrence of intermittency in the outer region derived by \citet{Brauckmann2013a}.
Furthermore, we demonstrated that the stability difference between the inner and outer partition, and therefore the fluctuation asymmetry, decrease with declining curvature and eventually disappear as $\eta\rightarrow 1$. As a consequence, also the detachment torque maximum vanishes for $\eta\gtrsim0.9$.

For this low-curvature TCF, we identified two new torque maxima, a broad one at $R_\Omega=0.2$ that exists for all studied $\Rey_S$ and a narrow one at $R_\Omega=0.02$ that emerges as $\Rey_S$ increases to $2\e{4}$. 
The broad torque maximum is accompanied by a strong vortical flow as evidenced by the large wall-normal velocity and can be reproduced in streamwise invariant simulations which also points to streamwise vortices causing a high angular momentum transport. 
On the other hand, the narrow torque maximum coincides with an effective mean flow where the angular momentum transport is maximized not due to large-amplitude vortices but because of correlated vortices resulting in an effective convective transport. The momentum transport in RPCF features the same characteristics and also shows the two maxima.

Moreover, our simulations at $\Rey_S=2\e{4}$ reveal that Nusselt numbers, profile slopes, and fluctuation amplitudes collapse well for $\eta\geq0.9$ when plotted as a function of $R_\Omega$ and agree with the corresponding quantities in RPCF. These results empirically demonstrate the convergence of TCF with small but finite curvature to RPCF. In this limit, the dependence on the system rotation parametrized by $R_\Omega$ becomes universal in both systems.
In addition, the angular momentum profiles in TCF show this universal behaviour since they collapse remarkably well for different radius ratios $\eta\geq0.5$ as long as no stabilised outer region occurs. Furthermore, we demonstrated that for the unstable flow ($R_\Omega>0$) the turbulent mean profiles in TCF and RPCF feature gradients in the centre that conform with neutral stability. 
Rayleigh's inviscid neutral stability criterion has an analogous counterpart in RPCF and implies that the central profile slopes $r\p_r\avg{\omega}$ and analogously $-\p_y\avg{u_x}$ approximate $-R_\Omega$. 
In the low-curvature TCF and in RPCF, the flattest $\omega$-profile (or $u_x$-profile) and the narrow torque maximum appear at almost the same $R_\Omega$ value as also found for smaller $\eta\leq0.909$ by \citet{Ostilla2014}.
However for $\eta=0.909$, these authors observed only one maximum in the torque for $\Rey_S\sim10^5$ to $10^6$ which appears at $R_\Omega=0.04$ close to our narrow maximum at $R_\Omega=0.02$ and 
shows the same signature in the profile gradient. This fact and the stronger growth of the narrow torque maximum with $\Rey_S$ suggest that it will outperform the broad maximum as the shear increases
further. Further simulations or experiments for low-curvature flows with $\eta>0.9$ and $\Rey_S$ beyond $2\e{4}$ could provide welcome tests of this prediction.

The study of the transition from TCF to RPCF has turned out to bring more changes in the flow structures
and the turbulent characteristics than anticipated. Perhaps the most intriguing observation is the
transition from a single maximum in torque to two maxima for $\eta\gtrsim 0.9$. We could here
characterize the two maxima by their relation to prominent features of the turbulent flow and momentum transport. While this provides some insights into their origin, a better physical understanding of the
maxima and an explanation for their remarkably universal position in $R_\Omega$ would be desirable.\\

%
%
We thank M. Avila, S. Grossmann, D. P. Lathrop, D. Lohse, S. Merbold, R. Ostilla, M. S. Paoletti for various stimulating discussions. We are grateful to Marc Avila for developing and providing the code used for our Taylor-Couette simulations. 
This work was supported in part by the Deutsche Forschungsgemeinschaft within the research group FOR 1182.
Most computations were carried out at the LOEWE-CSC in Frankfurt.

\bibliographystyle{jfm}

\bibliography{library}

\begin{thebibliography}{68}
\expandafter\ifx\csname natexlab\endcsname\relax\def\natexlab#1{#1}\fi

\bibitem[Ahlers {\em et~al.\/}(2012)Ahlers, Bodenschatz, Funfschilling,
  Grossmann, He, Lohse, Stevens \& Verzicco]{Ahlers2012}
{\sc Ahlers, G., Bodenschatz, E., Funfschilling, D., Grossmann, S., He, X.,
  Lohse, D., Stevens, R. J. A.~M. \& Verzicco, R.} 2012 {Logarithmic
  temperature profiles in turbulent Rayleigh-B\'{e}nard convection}. {\em Phys.
  Rev. Lett.\/} {\bf 109}, 114501.

\bibitem[Ahlers {\em et~al.\/}(2009)Ahlers, Grossmann \& Lohse]{Ahlers2009}
{\sc Ahlers, G., Grossmann, S. \& Lohse, D.} 2009 {Heat transfer and large
  scale dynamics in turbulent Rayleigh-B\'{e}nard convection}. {\em Rev. Mod.
  Phys.\/} {\bf 81}~(2), 503--537.

\bibitem[Andereck {\em et~al.\/}(1986)Andereck, Liu \& Swinney]{Andereck1986}
{\sc Andereck, C.~D., Liu, S.~S. \& Swinney, H.~L.} 1986 {Flow regimes in a
  circular Couette system with independently rotating cylinders}. {\em J. Fluid
  Mech.\/} {\bf 164}, 155--183.

\bibitem[Barri \& Andersson(2010)]{Barri2010}
{\sc Barri, M. \& Andersson, H.~I.} 2010 {Computer experiments on rapidly
  rotating plane Couette flow}. {\em Commun. Comput. Phys.\/} {\bf 7}~(4),
  683--717.

\bibitem[Bech \& Andersson(1997)]{Bech1997}
{\sc Bech, K.~H. \& Andersson, H.~I.} 1997 {Turbulent plane Couette flow
  subject to strong system rotation}. {\em J. Fluid Mech.\/} {\bf 347},
  289--314.

\bibitem[Bilson \& Bremhorst(2007)]{Bilson2007}
{\sc Bilson, M. \& Bremhorst, K.} 2007 {Direct numerical simulation of
  turbulent Taylor-Couette flow}. {\em J. Fluid Mech.\/} {\bf 579}, 227--270.

\bibitem[Bradshaw(1969)]{Bradshaw1969}
{\sc Bradshaw, P.} 1969 {The analogy between streamline curvature and buoyancy
  in turbulent shear flow}. {\em J. Fluid Mech.\/} {\bf 36}, 177--191.

\bibitem[Brauckmann \& Eckhardt(2013{\natexlab{{\em a\/}}})]{Brauckmann2013}
{\sc Brauckmann, H.~J. \& Eckhardt, B.} 2013{\natexlab{{\em a\/}}} {Direct
  numerical simulations of local and global torque in Taylor-Couette flow up to
  Re = 30 000}. {\em J. Fluid Mech.\/} {\bf 718}, 398--427.

\bibitem[Brauckmann \& Eckhardt(2013{\natexlab{{\em b\/}}})]{Brauckmann2013a}
{\sc Brauckmann, H.~J. \& Eckhardt, B.} 2013{\natexlab{{\em b\/}}}
  {Intermittent boundary layers and torque maxima in Taylor-Couette flow}. {\em
  Phys. Rev. E\/} {\bf 87}~(3), 033004.

\bibitem[Brethouwer {\em et~al.\/}(2012)Brethouwer, Duguet \&
  Schlatter]{Brethouwer2012}
{\sc Brethouwer, G., Duguet, Y. \& Schlatter, P.} 2012 {Turbulent-laminar
  coexistence in wall flows with Coriolis, buoyancy or Lorentz forces}. {\em J.
  Fluid Mech.\/} {\bf 704}, 137--172.

\bibitem[Brown \& Ahlers(2007)]{Brown2007}
{\sc Brown, E. \& Ahlers, G.} 2007 {Temperature gradients, and search for
  non-Boussinesq effects, in the interior of turbulent Rayleigh-B\'{e}nard
  convection}. {\em Europhys. Lett.\/} {\bf 80}~(1), 14001.

\bibitem[Burin \& Czarnocki(2012)]{Burin2012}
{\sc Burin, M.~J. \& Czarnocki, C.~J.} 2012 {Subcritical transition and spiral
  turbulence in circular Couette flow}. {\em J. Fluid Mech.\/} {\bf 709},
  106--122.

\bibitem[Burin {\em et~al.\/}(2010)Burin, Schartman \& Ji]{Burin2010}
{\sc Burin, M.~J., Schartman, E. \& Ji, H.} 2010 {Local measurements of
  turbulent angular momentum transport in circular Couette flow}. {\em Exp.
  Fluids\/} {\bf 48}, 763--769.

\bibitem[Chandrasekhar(1961)]{Chandrasekhar1961}
{\sc Chandrasekhar, S.} 1961 {\em {Hydrodynamic and Hydromagnetic
  Stability}\/}, 1st edn. Oxford: Clarendon Press.

\bibitem[Coles(1965)]{Coles1965}
{\sc Coles, D.} 1965 {Transition in circular Couette flow}. {\em J. Fluid
  Mech.\/} {\bf 21}, 385--425.

\bibitem[Coughlin \& Marcus(1996)]{Coughlin1996}
{\sc Coughlin, K. \& Marcus, P.~S.} 1996 {Turbulent bursts in Couette-Taylor
  flow}. {\em Phys. Rev. Lett.\/} {\bf 77}~(11), 2214--2217.

\bibitem[Daly {\em et~al.\/}(2014)Daly, Schneider, Schlatter \&
  Peake]{Daly2014}
{\sc Daly, C.~A., Schneider, T.~M., Schlatter, P. \& Peake, N.} 2014 {Secondary
  instability and tertiary states in rotating plane Couette flow}. {\em J.
  Fluid Mech.\/} {\bf 761}, 27--61.

\bibitem[Deguchi {\em et~al.\/}(2014)Deguchi, Meseguer \&
  Mellibovsky]{Deguchi2014}
{\sc Deguchi, K., Meseguer, A. \& Mellibovsky, F.} 2014 {Subcritical equilibria
  in Taylor-Couette flow}. {\em Phys. Rev. Lett.\/} {\bf 112}~(18), 184502.

\bibitem[Demay {\em et~al.\/}(1992)Demay, Iooss \& Laure]{Demay1992}
{\sc Demay, Y., Iooss, G. \& Laure, P.} 1992 {Wave patterns in the small gap
  Couette-Taylor problem}. {\em Eur. J. Mech. B/Fluids\/} {\bf 11}, 621--634.

\bibitem[Dong(2007)]{Dong2007}
{\sc Dong, S.} 2007 {Direct numerical simulation of turbulent Taylor-Couette
  flow}. {\em J. Fluid Mech.\/} {\bf 587}, 373--393.

\bibitem[Dubrulle(1993)]{Dubrulle1993}
{\sc Dubrulle, B.} 1993 {Differential rotation as a source of angular momentum
  transfer in the solar nebula}. {\em Icarus\/} {\bf 106}, 59--76.

\bibitem[Dubrulle {\em et~al.\/}(2005)Dubrulle, Dauchot, Daviaud, Longaretti,
  Richard \& Zahn]{Dubrulle2005}
{\sc Dubrulle, B., Dauchot, O., Daviaud, F., Longaretti, P.-Y., Richard, D. \&
  Zahn, J.-P.} 2005 {Stability and turbulent transport in Taylor-Couette flow
  from analysis of experimental data}. {\em Phys. Fluids\/} {\bf 17}~(9),
  095103.

\bibitem[Dubrulle \& Hersant(2002)]{Dubrulle2002}
{\sc Dubrulle, B. \& Hersant, F.} 2002 {Momentum transport and torque scaling
  in Taylor-Couette flow from an analogy with turbulent convection}. {\em Eur.
  Phys. J. B\/} {\bf 26}, 379--386.

\bibitem[Eckhardt {\em et~al.\/}(2000)Eckhardt, Grossmann \&
  Lohse]{Eckhardt2000}
{\sc Eckhardt, B., Grossmann, S. \& Lohse, D.} 2000 {Scaling of global momentum
  transport in Taylor-Couette and pipe flow}. {\em Eur. Phys. J. B\/} {\bf
  18}~(3), 541--544.

\bibitem[Eckhardt {\em et~al.\/}(2007{\natexlab{{\em a\/}}})Eckhardt, Grossmann
  \& Lohse]{Eckhardt2007a}
{\sc Eckhardt, B., Grossmann, S. \& Lohse, D.} 2007{\natexlab{{\em a\/}}}
  {Fluxes and energy dissipation in thermal convection and shear flows}. {\em
  Europhys. Lett.\/} {\bf 78}, 24001.

\bibitem[Eckhardt {\em et~al.\/}(2007{\natexlab{{\em b\/}}})Eckhardt, Grossmann
  \& Lohse]{Eckhardt2007}
{\sc Eckhardt, B., Grossmann, S. \& Lohse, D.} 2007{\natexlab{{\em b\/}}}
  {Torque scaling in turbulent Taylor-Couette flow between independently
  rotating cylinders}. {\em J. Fluid Mech.\/} {\bf 581}, 221--250.

\bibitem[Eckhardt \& Yao(1995)]{Eckhardt1995}
{\sc Eckhardt, B. \& Yao, D.} 1995 {Local stability analysis along Lagrangian
  paths}. {\em Chaos, Solitons \& Fractals\/} {\bf 5}~(11), 2073--2088.

\bibitem[Esser \& Grossmann(1996)]{Esser1996}
{\sc Esser, A. \& Grossmann, S.} 1996 {Analytic expression for Taylor-Couette
  stability boundary}. {\em Phys. Fluids\/} {\bf 8}~(7), 1814.

\bibitem[Faisst \& Eckhardt(2000)]{Faisst2000}
{\sc Faisst, H. \& Eckhardt, B.} 2000 {Transition from the Couette-Taylor
  system to the plane Couette system}. {\em Phys. Rev. E\/} {\bf 61},
  7227--7230.

\bibitem[Gibson(2012)]{channelflow}
{\sc Gibson, J.~F.} 2012 {{Channelflow}: {A} spectral {Navier-Stokes} simulator
  in {C}++}. {\em Tech. Rep.\/}. U. New Hampshire.

\bibitem[Gibson {\em et~al.\/}(2008)Gibson, Halcrow \&
  Cvitanovi\'{c}]{Gibson2008}
{\sc Gibson, J.~F., Halcrow, J. \& Cvitanovi\'{c}, P.} 2008 {Visualizing the
  geometry of state space in plane Couette flow}. {\em J. Fluid Mech.\/} {\bf
  611}, 107--130.

\bibitem[van Gils {\em et~al.\/}(2011)van Gils, Huisman, Bruggert, Sun \&
  Lohse]{VanGils2011}
{\sc van Gils, D. P.~M., Huisman, S.~G., Bruggert, G.-W., Sun, C. \& Lohse, D.}
  2011 {Torque scaling in turbulent Taylor-Couette flow with co- and
  counterrotating cylinders}. {\em Phys. Rev. Lett.\/} {\bf 106}, 024502.

\bibitem[van Gils {\em et~al.\/}(2012)van Gils, Huisman, Grossmann, Sun \&
  Lohse]{VanGils2012}
{\sc van Gils, D. P.~M., Huisman, S.~G., Grossmann, S., Sun, C. \& Lohse, D.}
  2012 {Optimal Taylor-Couette turbulence}. {\em J. Fluid Mech.\/} {\bf 706},
  118--149.

\bibitem[Hiwatashi {\em et~al.\/}(2007)Hiwatashi, Alfredsson, Tillmark \&
  Nagata]{Hiwatashi2007}
{\sc Hiwatashi, K., Alfredsson, P.~H., Tillmark, N. \& Nagata, M.} 2007
  {Experimental observations of instabilities in rotating plane Couette flow}.
  {\em Phys. Fluids\/} {\bf 19}~(4), 048103.

\bibitem[Huisman {\em et~al.\/}(2014)Huisman, van~der Veen, Sun \&
  Lohse]{Huisman2014}
{\sc Huisman, S.~G., van~der Veen, R. C.~A., Sun, C. \& Lohse, D.} 2014
  {Multiple states in highly turbulent Taylor-Couette flow}. {\em Nat.
  Commun.\/} {\bf 5}, 3820.

\bibitem[Kerr(1996)]{Kerr1996}
{\sc Kerr, RM} 1996 {Rayleigh number scaling in numerical convection}. {\em J.
  Fluid Mech.\/} {\bf 310}, 139.

\bibitem[Komminaho {\em et~al.\/}(1996)Komminaho, Lundbladh \&
  Johansson]{Komminaho1996}
{\sc Komminaho, J., Lundbladh, A. \& Johansson, A.~V.} 1996 {Very large
  structures in plane turbulent Couette flow}. {\em J. Fluid Mech.\/} {\bf
  320}, 259.

\bibitem[Landau \& Lifshitz(1987)]{Landau}
{\sc Landau, L.~D. \& Lifshitz, E.~M.} 1987 {\em {Fluid Mechanics}\/}, 2nd edn.
  New York: Pergamon Press.

\bibitem[Lathrop {\em et~al.\/}(1992)Lathrop, Fineberg \& Swinney]{Lathrop1992}
{\sc Lathrop, D.~P., Fineberg, J. \& Swinney, H.~L.} 1992 {Transition to
  shear-driven turbulence in Couette-Taylor flow}. {\em Phys. Rev. A\/} {\bf
  46}, 6390--6405.

\bibitem[Lewis \& Swinney(1999)]{Lewis1999}
{\sc Lewis, G.~S. \& Swinney, H.~L.} 1999 {Velocity structure functions,
  scaling, and transitions in high-Reynolds-number Couette-Taylor flow}. {\em
  Phys. Rev. E\/} {\bf 59}, 5457--67.

\bibitem[Lezius \& Johnston(1976)]{Lezius1976}
{\sc Lezius, D.~K. \& Johnston, J.~P.} 1976 {Roll-cell instabilities in
  rotating laminar and trubulent channel flows}. {\em J. Fluid Mech.\/} {\bf
  77}, 153--175.

\bibitem[Marcus(1984)]{Marcus1984a}
{\sc Marcus, P.~S.} 1984 {Simulation of Taylor-Couette flow. Part 1. Numerical
  methods and comparison with experiment}. {\em J. Fluid Mech.\/} {\bf 146},
  45--64.

\bibitem[Mart\'{\i}nez-Arias {\em et~al.\/}(2014)Mart\'{\i}nez-Arias, Peixinho,
  Crumeyrolle \& Mutabazi]{Martinez-Arias2014}
{\sc Mart\'{\i}nez-Arias, B., Peixinho, J., Crumeyrolle, O. \& Mutabazi, I.}
  2014 {Effect of the number of vortices on the torque scaling in
  Taylor-Couette flow}. {\em J. Fluid Mech.\/} {\bf 748}, 756--767.

\bibitem[Merbold {\em et~al.\/}(2013)Merbold, Brauckmann \&
  Egbers]{Merbold2013}
{\sc Merbold, S., Brauckmann, H.~J. \& Egbers, C.} 2013 {Torque measurements
  and numerical determination in differentially rotating wide gap
  Taylor-Couette flow}. {\em Phys. Rev. E\/} {\bf 87}, 023014.

\bibitem[Meseguer {\em et~al.\/}(2007)Meseguer, Avila, Mellibovsky \&
  Marques]{Meseguer2007}
{\sc Meseguer, A., Avila, M., Mellibovsky, F. \& Marques, F.} 2007 {Solenoidal
  spectral formulations for the computation of secondary flows in cylindrical
  and annular geometries}. {\em Eur. Phys. J. Spec. Top.\/} {\bf 146},
  249--259.

\bibitem[Nagata(1986)]{Nagata1986}
{\sc Nagata, M.} 1986 {Bifurcations in Couette flow between almost corotating
  cylinders}. {\em J. Fluid Mech.\/} {\bf 169}, 229.

\bibitem[Nagata(1990)]{Nagata1990}
{\sc Nagata, M.} 1990 {Three-dimensional finite-amplitude solutions in plane
  Couette flow: bifurcation from infinity}. {\em J. Fluid Mech.\/} {\bf 217},
  519.

\bibitem[Ostilla {\em et~al.\/}(2013)Ostilla, Stevens, Grossmann, Verzicco \&
  Lohse]{Ostilla2013}
{\sc Ostilla, R., Stevens, R. J. A.~M., Grossmann, S., Verzicco, R. \& Lohse,
  D.} 2013 {Optimal Taylor-Couette flow: direct numerical simulations}. {\em J.
  Fluid Mech.\/} {\bf 719}, 14--46.

\bibitem[Ostilla-M\'{o}nico {\em et~al.\/}(2014{\natexlab{{\em
  a\/}}})Ostilla-M\'{o}nico, Huisman, Jannink, {Van Gils}, Verzicco, Grossmann,
  Sun \& Lohse]{Ostilla2014}
{\sc Ostilla-M\'{o}nico, R., Huisman, S.~G., Jannink, T. J.~G., {Van Gils}, D.
  P.~M., Verzicco, R., Grossmann, S., Sun, C. \& Lohse, D.} 2014{\natexlab{{\em
  a\/}}} {Optimal Taylor-Couette flow: radius ratio dependence}. {\em J. Fluid
  Mech.\/} {\bf 747}, 1--29.

\bibitem[Ostilla-M\'{o}nico {\em et~al.\/}(2014{\natexlab{{\em
  b\/}}})Ostilla-M\'{o}nico, van~der Poel, Verzicco, Grossmann \&
  Lohse]{Ostilla-Monico2014b}
{\sc Ostilla-M\'{o}nico, R., van~der Poel, E.~P., Verzicco, R., Grossmann, S.
  \& Lohse, D.} 2014{\natexlab{{\em b\/}}} {Exploring the phase diagram of
  fully turbulent Taylor-Couette flow}. {\em J. Fluid Mech.\/} {\bf 761},
  1--26.

\bibitem[Ostilla-M\'{o}nico {\em et~al.\/}(2015)Ostilla-M\'{o}nico, Verzicco \&
  Lohse]{Ostilla-Monico2015a}
{\sc Ostilla-M\'{o}nico, R., Verzicco, R. \& Lohse, D.} 2015 {Effects of the
  computational domain size on DNS of Taylor-Couette turbulence}. {\em Submitt.
  to JFM\/} p. arXiv:1411.3826v1.

\bibitem[Paoletti {\em et~al.\/}(2012)Paoletti, van Gils, Dubrulle, Sun, Lohse
  \& Lathrop]{Paoletti2012}
{\sc Paoletti, M.~S., van Gils, D. P.~M., Dubrulle, B., Sun, C., Lohse, D. \&
  Lathrop, D.~P.} 2012 {Angular momentum transport and turbulence in laboratory
  models of Keplerian flows}. {\em Astron. Astrophys.\/} pp. 1--11.

\bibitem[Paoletti \& Lathrop(2011)]{Paoletti2011}
{\sc Paoletti, M.~S. \& Lathrop, D.~P.} 2011 {Angular momentum transport in
  turbulent flow between independently rotating cylinders}. {\em Phys. Rev.
  Lett.\/} {\bf 106}, 024501.

\bibitem[Pope(2000)]{Pope}
{\sc Pope, S.} 2000 {\em {Turbulent flows}\/}. Cambridge University Press.

\bibitem[Ravelet {\em et~al.\/}(2010)Ravelet, Delfos \&
  Westerweel]{Ravelet2010}
{\sc Ravelet, F., Delfos, R. \& Westerweel, J.} 2010 {Influence of global
  rotation and Reynolds number on the large-scale features of a turbulent
  Taylor-Couette flow}. {\em Phys. Fluids\/} {\bf 22}~(5), 055103.

\bibitem[Rayleigh(1917)]{Rayleigh1917}
{\sc Rayleigh, L.} 1917 {On the dynamics of revolving fluids}. {\em Proc. R.
  Soc. London A\/} {\bf 93}~(648), 148--154.

\bibitem[Salewski \& Eckhardt(2015)]{Salewski2015}
{\sc Salewski, M. \& Eckhardt, B.} 2015 {Turbulent states in plane Couette flow
  with rotation}. {\em Phys. Fluids\/} {\bf 27}~(4), 045109.

\bibitem[Smith \& Townsend(1982)]{Smith1982}
{\sc Smith, G.~P. \& Townsend, A.~A.} 1982 {Turbulent Couette flow between
  concentric cylinders at large Taylor numbers}. {\em J. Fluid Mech.\/} {\bf
  123}, 187--217.

\bibitem[Suryadi {\em et~al.\/}(2014)Suryadi, Segalini \&
  Alfredsson]{Suryadi2014}
{\sc Suryadi, A., Segalini, A. \& Alfredsson, P.~H.} 2014 {Zero absolute
  vorticity: Insight from experiments in rotating laminar plane Couette flow}.
  {\em Phys. Rev. E\/} {\bf 89}~(3), 033003.

\bibitem[Taylor(1923)]{Taylor1923}
{\sc Taylor, G.~I.} 1923 {Stability of a viscous liquid contained between two
  rotating cylinders}. {\em Philos. Trans. R. Soc. London, Ser. A\/} {\bf 223},
  289--343.

\bibitem[Taylor(1935)]{Taylor1935}
{\sc Taylor, G.~I.} 1935 {Distribution of velocity and temperature between
  concentric rotating cylinders}. {\em Proc. R. Soc. London A\/} {\bf 151},
  494--512.

\bibitem[Taylor(1936)]{Taylor1936a}
{\sc Taylor, G.~I.} 1936 {Fluid friction between rotating cylinders. I. Torque
  measurements}. {\em Proc. R. Soc. London A\/} {\bf 157}, 546--564.

\bibitem[Tilgner {\em et~al.\/}(1993)Tilgner, Belmonte \&
  Libchaber]{Tilgner1993}
{\sc Tilgner, A., Belmonte, A. \& Libchaber, A.} 1993 {Temperature and velocity
  profiles of turbulent convection in water}. {\em Phys. Rev. E\/} {\bf
  47}~(4), R2253.

\bibitem[Tritton(1992)]{Tritton1992}
{\sc Tritton, D.~J.} 1992 {Stabilization and destabilization of turbulent shear
  flow in a rotating fluid}. {\em J. Fluid Mech.\/} {\bf 241}, 503.

\bibitem[Tsukahara(2011)]{Tsukahara2011}
{\sc Tsukahara, T.} 2011 {Structures and turbulent statistics in a rotating
  plane Couette flow}. {\em J. Phys. Conf. Ser.\/} {\bf 318}~(2), 022024.

\bibitem[Tsukahara {\em et~al.\/}(2010)Tsukahara, Tillmark \&
  Alfredsson]{Tsukahara2010}
{\sc Tsukahara, T., Tillmark, N. \& Alfredsson, P.~H.} 2010 {Flow regimes in a
  plane Couette flow with system rotation}. {\em J. Fluid Mech.\/} {\bf 648},
  5--33.

\bibitem[Wattendorf(1935)]{Wattendorf1935}
{\sc Wattendorf, F.~L.} 1935 {A study of the effect of curvature on fully
  developed turbulent flow}. {\em Proc. R. Soc. London A\/} {\bf 148},
  565--598.

\bibitem[Wendt(1933)]{Wendt1933}
{\sc Wendt, F.} 1933 {Turbulente Str\"{o}mungen zwischen zwei rotierenden
  konaxialen Zylindern}. {\em Ing. Arch.\/} {\bf 4}, 577--595.

\end{thebibliography}

\end{document}